# Spatial Variation and Correlation of Spin Properties in Organic Light-Emitting Diodes

W. J. Pappas[1], R. Geng[1], A. Mena[1], A. Baldacchino[1], A. Asadpoordarvish[1], D. R. McCamey[1,*]

[1]ARC Centre of Excellence in Exciton Science, School of Physics, UNSW Sydney, NSW 2052, Australia

*Corresponding author:   dane.mccamey@unsw.edu.au

**Abstract**

Devices which exploit the quantum properties of materials are widespread, with quantum information processors and quantum sensors showing significant progress. Organic devices offer interesting opportunities for quantum technologies owing to their engineerable spin properties, with spintronic operation and spin resonance magnetic-field sensing demonstrated in research grade devices, as well as proven compatibility with large scale fabrication techniques. Yet several important challenges remain as we move toward scaling these proof-of-principle quantum devices to larger integrated logic systems or spatially smaller sensing elements, particularly those associated with the variation of quantum properties both within and between devices. Here, spatially resolved magnetoluminescence is used to provide the first two-dimensional map of a spin property – the Overhauser field – in an organic light emitting diode. We find intra-device variabilities exceeding 20% while spatially correlated behaviour is exhibited on lengths beyond 7 μm, similar in size to pixels in state-of-the-art AMOLED arrays, which has implications for the reproducibility and integration of organic quantum devices.

**Introduction – Variation of Spin-Properties in Organic Semiconductors**

Spin is a versatile resource for technology. It can form the basis of information processing or sensing elements[1–5], and its quantum nature provides advantages over other approaches[6]. To be most effective, spins must reside in materials which support long coherence and lifetimes, whilst retaining the ability to be interrogated[7]. Organic electronic materials provide a platform for such technologies[8,9] – they generally have weak spin-orbit coupling[10,11], support long lifetimes[12–14], allow for transduction between electrical and optical signals[15,16], and can be engineered using the many tools of synthetic chemistry[17–19]. Long spin diffusion lengths also support their use in spintronic logic devices[20,21], and the sizeable modulations in optoelectronic processes, such as conductance and luminescence, under the application of relatively weak magnetic fields point toward their application as quantum sensors[22,23]. Their fabrication flexibility and industrial use in commercial displays[24,25] underlies recent advances towards the miniaturisation and integration of quantum technologies based on these materials[26–28].

For effective integration or miniaturisation, reproducibility is crucial, and this extends to the properties of materials which impact spin-dependent processes. In this work we aim to measure the variation of a critical property common to organic materials, the Overhauser field, which arises from the hyperfine coupling of charge carrier spins to the bath of randomly oriented nuclear spins in the molecular environment[29]. The Overhauser field underlies a range of phenomena observed in organic devices, including magnetoresistive and magnetic resonant effects in organic light emitting diodes (OLEDs)[30–35], both of which have been proposed as a mechanism to enable magnetic field sensing[22,32]. To date, the Overhauser field has been treated as a bulk property of the device, characterised by a single value which describes its impact[32,36]. If this is not a good assumption, it will present a significant challenge as we move toward high spatial resolution of magnetic fields[37,38] in organic devices[39,40]. We will demonstrate below that the Overhauser field indeed shows substantial intra-device variation by spatially resolving the magnetoluminescence effect exhibited by a copolymer Super Yellow poly(phenylene-vinylene) (SY-PPV) OLED. As the Overhauser field is central to a wide range of spin-enabled functionality in organic devices, this variation presents a fundamental challenge for efforts to miniaturise and integrate such devices.

**Magnetoelectroluminescence in OLEDs**

We employ a stable and widely investigated active material – SY-PPV – to characterise the spatial variation of the Overhauser field in a simple vertical structure OLED (Figure 1a) and extract the length over which this property is correlated. In optoelectronic devices with bipolar charge transport, weakly coulombically bound electron-hole pairs (polaron pairs) form in the organic layer as an intermediate step to exciton formation[41]. The overall spin permutation symmetry of these pairs are modified by the application of external magnetic fields which then modulates their spin-sensitive optoelectronic processes[34,35]. That is, shifts in the concentrations of singlet- and triplet-character dominated polaron pairs under the application of static magnetic fields leads to changes in both the device conductivity and electroluminescence[42]. The main component of this change is due to the interplay between the external magnetic field and local Overhauser fields stemming from the hyperfine interaction of polarons with the spin bearing nuclei of molecules on which they reside[43,44]. In π-conjugated polymers, local variations in the strengths of these spin interactions arise from the complex spatial profile of polarons which delocalise over several monomer units[19,45]. Each of the nuclear spins encompassed within a given wavefunction couples to the spin of that charge carrier, leading to the precession of polaron spins about an effective (Overhauser) field[46], $\overrightarrow{B_{hf}}$ (Figure 1b). As both the delocalisation length of polaron wavefunctions and the orientations of individual nuclear spins vary[47], substantial variation in the Overhauser fields throughout the material results in the low-field magnetic effects observed in most organic optoelectronic devices[40,48].

The magnetoelectroluminescence (MEL) effect is used to investigate the spin-sensitive luminescence of OLEDs and is intrinsically related to the hyperfine interaction. Due to local hyperfine disorder, the Overhauser fields experienced by each carrier within a spin-pair are randomly oriented and possess distinct magnitudes[49]. As a consequence, the pair partners precess at different rates ($|\Delta\omega_{hf}|$) and are aligned along distinct axes, causing the overall spin-permutation symmetry to evolve between more singlet- and more triplet-like configurations[46]. The introduction of a static external magnetic field $\overrightarrow{B_{ext}}$ produces a new effective field ($\overrightarrow{B_{eff}} = \overrightarrow{B_{hf}} + \overrightarrow{B_{ext}}$) which modifies charge carrier precession. Under the application of small external fields $B_{ext} \sim B_{hf}$, spin mixing between the singlet and triplet manifolds remains dominated by the random distribution in Overhauser fields. However, at larger applied fields $B_{ext} \gg B_{hf}$, the precession disparity between charge carriers becomes insignificant and spin mixing is suppressed[43] (Figure 1c). As the rates of dissociation and recombination differ for singlet- and triplet-like polaron pairs, the stabilisation of spin configurations results in changes to the steady-state electroluminescence of the OLED[50,51].

Monolithic measurements of MEL for several current densities are displayed in Figure 1d. The observed growth in electroluminescence closely follows typical lineshapes reported in literature[36,52] and are well fit with a function of the form $\Delta EL(B_{ext})/EL = AB_{ext}^2/(|B_{ext}| + B_0)^2$, where A is the magnitude of the effect and $B_0$ is the characteristic width. To obtain a measure of the variation in Overhauser fields, we use the half-width at half-maximum of MEL curves ($B_{1/2} = (1 + \sqrt{2})B_0$) which has been experimentally demonstrated to scale with the hyperfine interaction strength, $a_{HF}$[19,53]. Interestingly, A exhibits a monotonic decrease with increasing current density which has been previously attributed to the competition between spin mixing and operation bias-dependent dissociation of polaron pairs[54]. Conversely, the bulk $B_{1/2}$ parameter shows no bias dependence which we independently verify using magnetic resonance techniques that provide a direct measure of the average Overhauser field distribution (Figure S19). While useful for characterising the bulk spin properties, monolithic measures of magnetic field effects supress vital information about the intra-device variation and spatial distribution of underlying quantum properties, which we now seek to investigate by spatially resolving MEL.

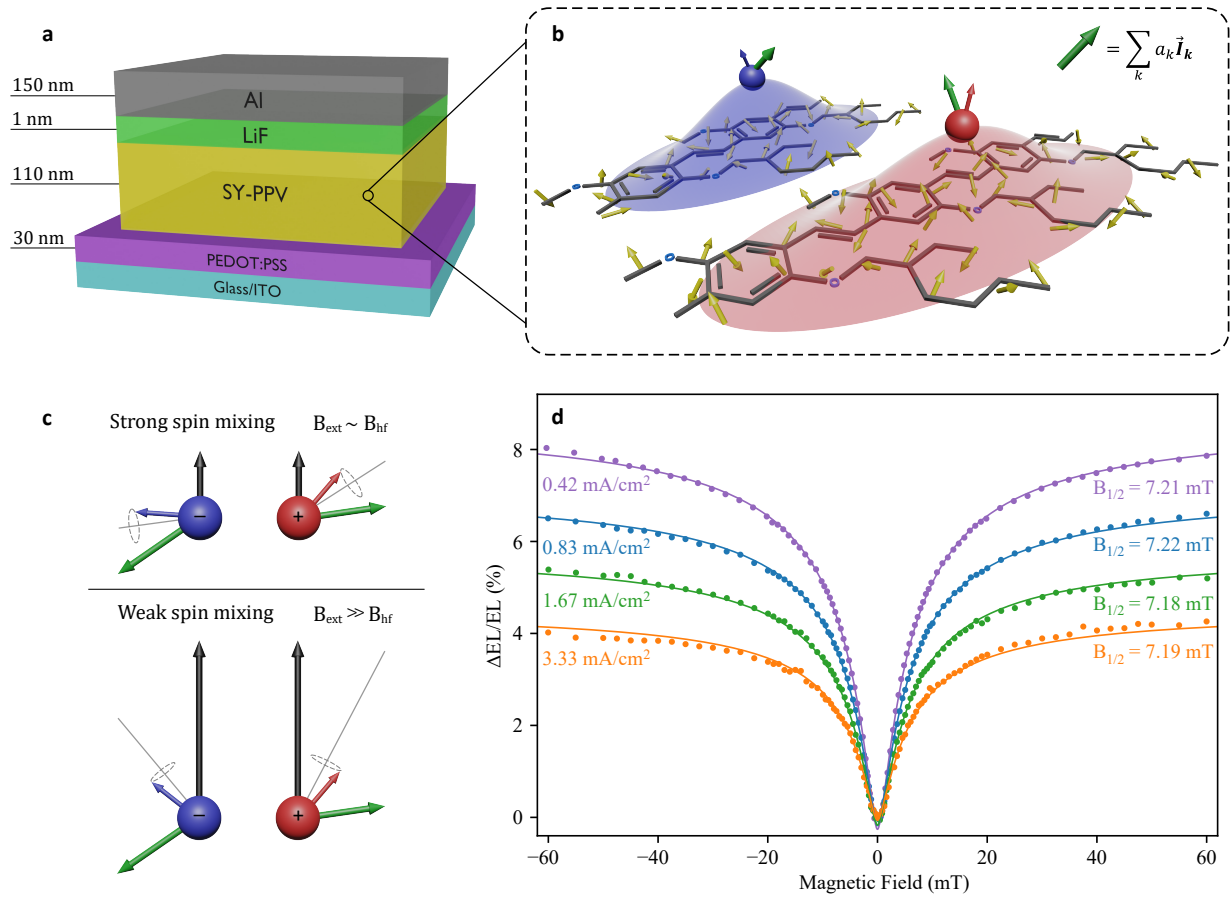

**Figure 1**: (a) Device architecture of the vertical structure SY-PPV OLED. (b) Illustration of the hyperfine interaction in the organic polymer. The polaron wavefunctions of electron (blue) and hole (red) charge carriers delocalise over the spin-bearing nuclei found in their molecular environments and lead to the precession of carrier spins about local Overhauser fields (green arrows). (c) Charge carrier spin orientations in an electron-hole pair under the application of a weak and strong external magnetic field relative to the absolute difference in Overhauser fields, $|\Delta B_{hf}|$. Larger fields minimise the disparity in Larmor precessions and act to suppress spin mixing. (d) Monolithic MEL measurements taken as the average EL over several magnetic field values for a large area of the device at several current densities. Bulk MEL lineshapes are fit with the specific non-Lorentzian function as described in the main text. Note that the magnitude of the effect diminishes with increasing current density, while the half-width at half-maximum ($B_{1/2}$) exhibits no current dependence.

**Spatially Resolved Magnetoelectroluminescence**

Spatial resolution of MEL was enabled by incorporating optical microscopy into a traditional MEL setup (Figure 2a). The OLED, positioned in a uniform external magnetic field, was centered below a microscope objective which imaged the device onto a scientific CMOS detector. The limit of resolution for our optical system was constrained by the CMOS sensor at ~1 μm under 20x magnification, which lies above the lateral resolution of the objective at ~700 nm for the peak spectral response of the OLED. Furthermore, a resonator based on designs used in similar spin-sensing systems[55] was incorporated to enable magnetic resonance to be performed. Details of the experimental apparatus can be found in the supplementary information (SI).

The magnetic-field induced change in OLED electroluminescence was imaged as a function of external magnetic field. At each field strength, the change in electroluminescence intensity relative to the zero-field intensity (ΔEL/EL) was calculated for each pixel in the CMOS sensor array. Figure 2b displays a ΔEL/EL map at 10 mT which shows microscopic variations in the ΔEL/EL that are significant and spatially-correlated on scales larger than the optical resolution of our setup. These structures exhibit a high degree of correlation between successive field measurements (Figure S12) and exceed the sizes of correlated electroluminescence seen previously in OLEDs[56], indicating that the magnetic-field-induced correlations are not due to the optical properties of the device. Similar maps captured at several fields of increasing strength are shown in Figure 2c. The ΔEL/EL for several small regions of these maps is shown in Figure 2d (4 x 4 pixels, identified by the vertical lines in 2c) as a function of the external magnetic field. Fits to this spatially resolved MEL data reveals substantial variations in the amplitudes and half-widths, while also uncovering additional variation between the two sides of the curve (which are fit independently).

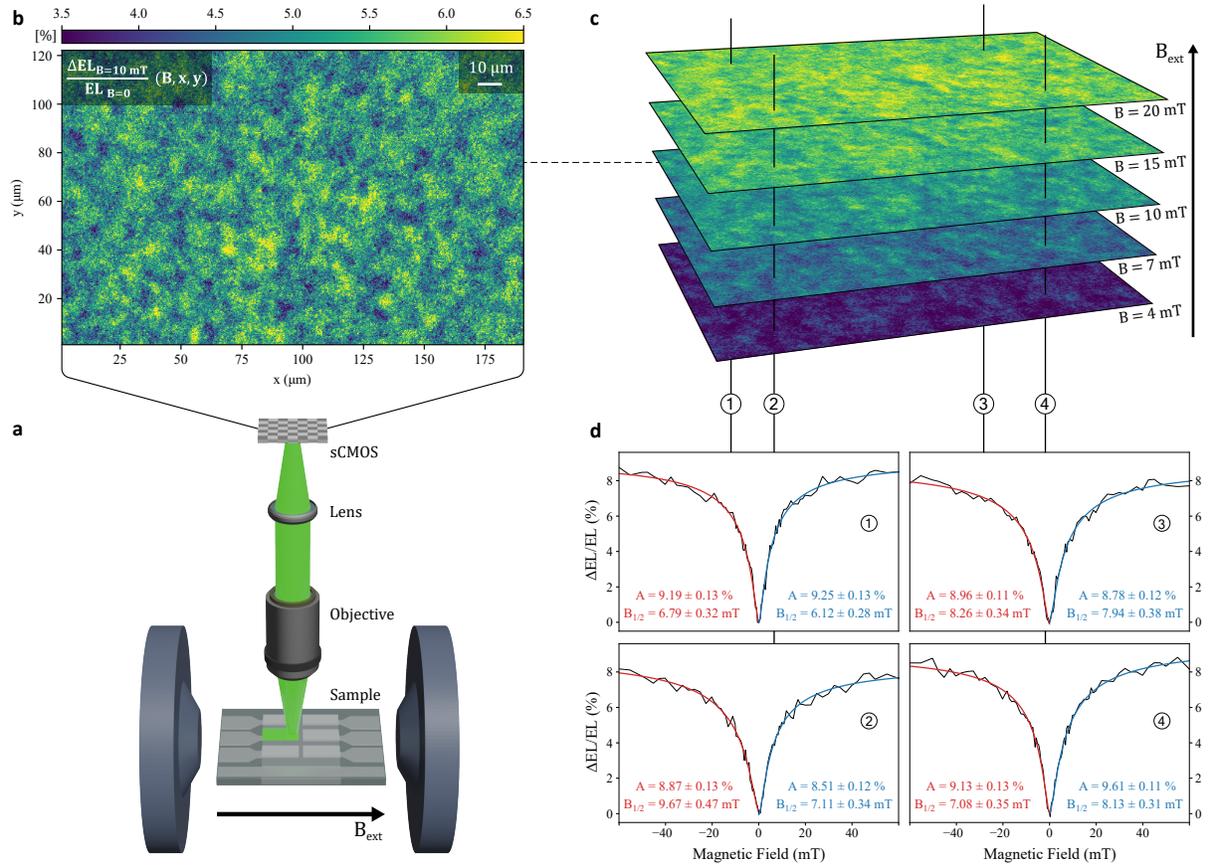

**Figure 2**: (a) A simplified illustration of the experimental setup with a focus on the optical system. An OLED is positioned centrally between a homogeneous magnetic field which is swept over both field polarities. The resulting field-dependent electroluminescence is collected under 20x magnification by an infinity-corrected objective and is subsequently refocused onto the sensor of a sCMOS camera. (b) An image of the electroluminescence intensity change at 10 mT relative to the zero-field frame (ΔEL/EL). Spatially correlated structure in deviations to the mean ΔEL/EL of up to 1.5% are observed and evolve over successive field measurements. (c) Selected ΔEL/EL maps at several magnetic fields are displayed vertically with increasing magnetic field strength. Vertical lines cut through the maps at four chosen spatial coordinates which form part of the MEL curves over 1.7 μm² regions. (d) The corresponding spatially resolved MEL curves for the entire magnetic field sweep at the chosen coordinates. We note that in addition to substantial spatial variation in the amplitudes and half-widths, MEL curves evolve temporally between positive and negative components of the field sweep. Both the spatial and temporal variations are hidden in monolithic measurements which reflect the bulk spin properties of the device.

To discern the spatial distributions of spin properties across the device, single pixel MEL curves are fit using the same function for each of the 625 x 400 pixels within an ~190 μm x 125 μm region of the device. Fit quality is high for each of the resulting 250,000 spatially resolved MEL curves with a mean $R^2$ ~ 0.97 (Figure S9). To minimise the impact of temporal changes in the Overhauser fields, we fit curves to only the positive or negative field data in each sweep. The fitting parameters are then extracted from each curve, allowing us to generate amplitude and half-width spatial maps. These parameter maps provide an excellent measure of the spatial variation and correlation of MEL (and subsequently the Overhauser fields) across the device.

**Spatial Variation and Correlation of Spin Properties**

Spatial maps are constructed from fits to the negative field side of the spatially resolved MEL and reveal the spatial distributions of amplitude (A) and half-width ($B_{1/2}$) parameters. Figure 3a and 3b illustrate this for A and $B_{1/2}$ respectively in the OLED operating at a current density of 0.42 mA/cm$^2$. Widespread correlated structure on the order of several micrometers is observed, with the $B_{1/2}$ parameter revealing that the Overhauser field varies substantially across the device. We now turn to a more detailed analysis of the variation and correlation which is extracted from parameter maps.

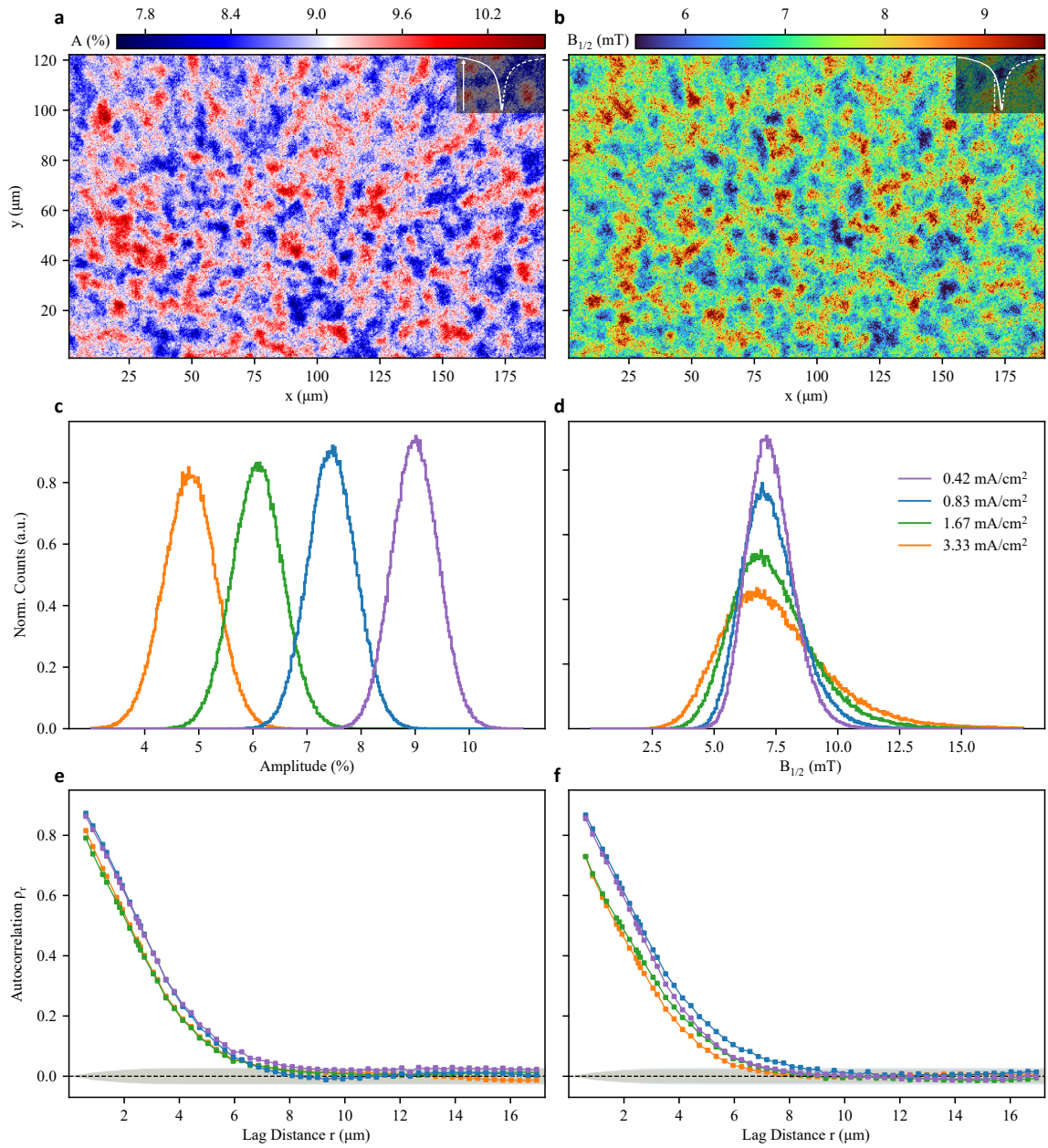

**Figure 3**: The spatial distribution, variation and autocorrelation of the amplitude, A (left), and half-width, $B_{1/2}$ (right), parameters extracted from the spatially resolved MEL fits. (a, b) The two-dimensional parameter maps generated from MEL resolved at an OLED current density of 0.42 mA/cm². Clear structure in the spatial distribution of parameters is observed, revealing regions of high spatial correlation in parameter space. The MEL fit parameterisation is overlayed (top right) on each map. (c, d) Area-normalised histograms quantify the variation and reveal the distribution of each parameter at four separate current densities. The mean amplitude, $A_{mean}$, exhibits clear dependence on the current density, whereas that of the half-width does not. On the other hand, the variability of $B_{1/2}$ distributions grows substantially with increasing current density, while that of A remains relatively unaffected. (e, f) The spatial autocorrelation as a function of lag distance (correlograms). A positive spatial autocorrelation indicates that parameters have a high likelihood of being similarly valued at the corresponding separation distances. Correlograms for both A and $B_{1/2}$ exhibit a statistically significant positive correlation for lengths of up to ~ 7 μm.

The variation of MEL can be seen through histograms containing the A and $B_{1/2}$ parameters extracted using the fitting process outlined above and are shown for several current densities in Figure 3c and 3d, respectively. These variations are well described by normal distributions (for fit parameters see Table S1). In agreement with bulk MEL fits, the mean amplitude ($A_{mean}$) decreases with increasing current density, whereas that of $B_{1/2}$ remains unchanged. However, the spatially resolved data now allows us to quantify the intra-device variation of these properties – A displays a similar dispersion ($A_{FWHM}$ = 1.0 – 1.1%) across all operating biases, translating to a minimum variability of 11.2% ($A_{FWHM}/A_{mean}$) at the lowest current density of 0.42 mA/cm$^2$, where the minimum variability of $B_{1/2}$ is 29.2%. Both sets of distributions exhibit larger variabilities which grow with increasing current density, displaying maximum variations of 23.5% and 63.0% for A and $B_{1/2}$, respectively. Interestingly, while a reduction in the signal-to-noise of MEL curves is observed with a larger operating bias (despite maintaining similar detection intensities of ~4.5 x 10$^4$ counts), fit uncertainties are not associated with the extremities of these distributions, in particular with the heavy-tailed lineshapes emerging in the $B_{1/2}$ distributions (Figure S10). The general broadening of these histograms instead points to the modification in Overhauser field distributions felt by polaron pairs and can likely be attributed to changes in the characteristic size of polaron wavefunctions when a bias is applied, rather than to a change in the material properties. As a larger bias allows polaron pairs with smaller binding energies to dissociate more rapidly[41], a reduction in pairs available for spin mixing can lead to a reduced MEL amplitude while resulting in a greater contribution to the effect from pairs that are more highly localised. This change in distribution is not discernible in monolithic measures, where no alteration in the bulk $B_{1/2}$ value (average Overhauser field) is detected, demonstrating a key advantage of this approach.

There are several approaches to quantify the spatial correlation of properties we extract from the spatially resolved MEL. Here, we apply spatial autocorrelation techniques using Moran's I (see methods for details) which provides a global measure of the likelihood of correlation between two points separated by a distance r.

The correlograms in Figure 3e and 3f show the spatial autocorrelation for A and $B_{1/2}$, respectively. At large distances, the correlation tends to zero, implying that no long-range order in the spatial distribution of either parameter exists. However, at shorter distances, the autocorrelation is positive, exceeding 0.5 at distances of r < 2 μm, indicating that regions separated on these length scales are significantly more likely to have similar values of A and $B_{1/2}$ than could be expected from a random distribution. This lies well above the resolution limit of our optical system. Moreover, correlograms are not observed to appreciably depend on the operating bias, with the small reduction in short-distance spatial autocorrelation associated with an increased variability of parameter distributions at higher current densities. In addition, the correlograms converge towards zero at similar distances, indicating that the dispersity of correlated Overhauser fields remains unaffected by the organic layer charge density. We calculate confidence bands using 95% confidence intervals and use this as the threshold to determine the statistical significance of correlation lengths for each parameter. At the lowest current density, A and $B_{1/2}$ exhibit significant correlation lengths for distances up to 8.4 μm and 7.5 μm, respectively.

We now consider the similarities between parameter maps. While correlated structures for A and $B_{1/2}$ are comparable in size, their local spatial distributions vary. We measure the similarity in distributions by calculating the cross-correlation between parameters for 10,000 separate 1.7 μm² (4 x 4 pixel) regions in the spatial maps. A Pearson coefficient of 0.16 and 0.24 ($p < 10^{-5}$) for negative and positive field components of MEL is found at 0.42 mA/cm². A positive correlation is seen for all current densities, indicating a significant inter-parameter relationship exists (Figure S17). Our method of ascertaining the cross-correlation between MEL parameters offers several distinct advantages over those used in previous efforts to study this effect[57], as here statistics are drawn from a single measurement performed on only one device.

To ensure reproducibility, spatial correlations of individual parameters between independent fits to positive and negative field components of MEL are performed. These show statistically significant Pearson autocorrelations of the spatially resolved A and $B_{1/2}$ parameters of 0.272 and 0.069 (with $p < 10^{-5}$) respectively, demonstrating the reproducibility of the approach, but also pointing to slow changes in the spatial distribution of the Overhauser field. The intra-device variation and spatial correlation of these maps however are highly consistent for each field component of MEL as they reflect global measures of these properties (Figures S15). We have also performed a set of faster measurements with a correspondingly lower signal-to-noise ratio that produce substantially higher correlations of A (0.71, $p < 10^{-5}$) and $B_{1/2}$ (0.24, $p < 10^{-5}$) between independent fits to each MEL component (Figure S16). This indicates that that there is likely a temporal effect in MEL that leads to increased intra-device variation, which will be addressed in future work. Nonetheless, identifying the existence of a temporal evolution in Overhauser fields highlights the power of resolving spatial information of quantum properties in organic semiconductors that would otherwise remain hidden when probing the device monolithically.

**Impacts of Overhauser Field Variation on Spin-Reliant Applications**

We now make an inference about the variation in quantum interactions which underlies the variation we observe in MEL and consider the impact of this on spin-based organic devices. Previous work has used deuteration in conjunction with magnetic resonance techniques to convincingly demonstrate that the half-width of the MEL curve is directly related to the strength of the Overhauser field[19,53]. Both electrical and optical detection methods of magnetic resonance exhibit a similar inhomogeneous broadening due to hyperfine disorder throughout the device (Figure S19), providing a direct measure of the global Overhauser field distributions independently experienced by the two polaron species comprising the spin-pairs (for further details refer to the SI). This technique can be effectively utilised for applications in spin sensing, particularly through OLED magnetometry[22,28]. However, the substantial intra-device variation of the characteristic Overhauser field strength as revealed by the spatially resolved MEL implies that the widths of such distributions vary by a similar amount (Figure 4). This is consequential for the performance and reproducibility for organic semiconductors which undergo miniaturisation and whose sensitivity and range depend on this hyperfine-induced broadening.

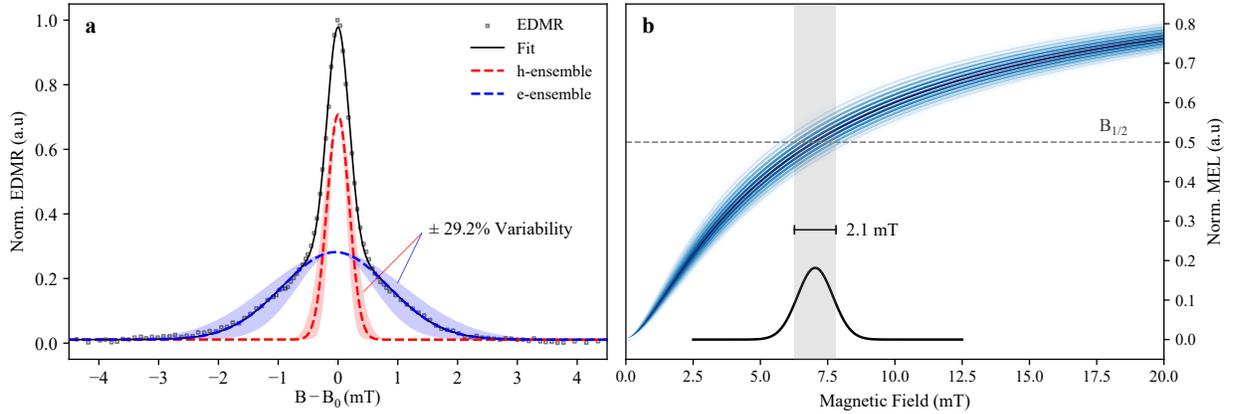

**Figure 4**: (a) The magnetic resonance spectrum (EDMR) dissected into its constituent components, the broadening of which reflect the global Overhauser field distributions experienced by electron and hole polaron ensembles in the organic layer. The black line shows the double Gaussian fit to EDMR data, while the blue and red dashed lines correspond to the resonance components stemming from electron and hole ensembles, respectively. The shaded regions illustrate the effect of a 29.2% variability in the linewidths for each component (FWHM of 2.09 mT and 0.43 mT for electron and hole ensembles, respectively) and reflect the intra-device spatial variation of Overhauser field distributions. (b) The $B_{1/2}$ distribution (black line) with a variability of 29.2% (FWHM 2.10 mT), as measured from the spatially resolved MEL, overlayed on a series of normalised MEL curves generated using $B_{1/2}$ values from that distribution. Both the EDMR and MEL measurements were performed on the same device and at a current density of 0.42 mA/cm$^2$.

Despite providing the ability to probe Overhauser distributions directly, electrically detected magnetic resonance (EDMR) is blind to the spatial information which would allow one to probe local variations to these distributions, while its optical counterpart (ODMR) is limited to extensive averaging and lock-in techniques due to the weak signal produced under the resonance condition[23]. In contrast, MEL exhibits much larger magnetically induced changes in luminosity whose half-width can be leveraged to extract the spatially varying hyperfine strength. However, MEL relies on overcoming the interplay between spins in their pairs – that is, the angle between spin alignment as well as their effective Larmor frequency. This results in a single normal distribution of effective Overhauser field strengths, the variability of which exceeds 20%. This variation manifests itself as a series of intra-device MEL curves with varying widths, characterised by the $B_{1/2}$ distribution as shown in Figure 4b. One can then extrapolate this finding to the magnetic resonance curves, where the shaded regions around each component of the Overhauser distribution in Figure 4a illustrates the effects of a 29.2% variation of this property throughout the device.

**Conclusion**

We have measured the spatial variation and correlation of the amplitude and width of magnetoelectroluminescence in an OLED. We find that these parameters vary by at least 11% and 29% respectively across the device and are correlated over distances of approximately 7 μm. As widths of spatially resolved MEL curves are related to the Overhauser fields experienced by charge carriers, we infer that this variation arises due to variations in the distribution of Overhauser fields in different regions of the device. We also find that an applied bias serves to vary the distribution of Overhauser fields, likely by changing the characteristic size of polarons contributing to MEL.

The results described above have implications in a wide range of applications – it places limits on the precision for spatially resolved magnetic field effect sensors[39], it provides a limit on the size of regions in which cooperative phenomena (such as the spin-Dicke effect[53]) are likely to occur on, and it points to materials challenges for the miniaturisation of spin-based devices below current state-of-the-art AMOLED pixel sizes. Further development of this experimental approach will also enable the investigation of spatiotemporal nuclear-spin dynamics in a range of organic systems utilised in quantum-based devices.

**Methods**

**Device Fabrication**

The OLEDs used in this study were fabricated in the following vertical structure: glass/ITO (~115 nm) / PEDOT:PSS (~32 nm)/SY-PPV (~110 nm)/LiF (~1 nm)/Al (~150 nm), where PEDOT:PSS is poly(3,4-ethylenedioxythiophene) polystyrene sulfonate and SY-PPV is super yellow light emitting PPV copolymer. The pre-patterned ITO substrates (Ossila S101) allow for six independent devices to be fabricated each with an active area of 1.5 mm x 4 mm. ITO Substrates were washed sequentially with Decon-90, distilled water, acetone, and propan-2-ol for 10 minutes each in an ultrasonic bath at 24°C (VWR Ultrasonic Cleaner USC-THD). They were then placed in a vacuum drying oven (DZ-1BCII) at 120°C for at least 12 hours.

After reaching ambient temperature in a fume cupboard, the PEDOT:PSS layer (Heraeus, Al 4083) was deposited via spin coating at 3000 rpm for 1 minute and annealed for 12 hours at 120°C. The substrates were then transferred to a glove box ($O_2$ <0.1 ppm, $H_2O$ < 0.1 ppm) where the SY-PPV layer was deposited via spin coating at 1500 rpm for 2 minutes. The SY-PPV solution was prepared in toluene with a concentration of 6 mg/ml, and then stirred with a magnetic bar at 60°C for 2 days before use. The SY-PPV solution was filtered using a PTFE syringe filters (0.45 μm pore size) before spin coating to remove any aggregates, resulting in a uniform thin-film with few defects. The samples were then transferred to a high vacuum chamber (< $10^{-8}$ mbar) for the inorganic layer depositions. After fabrication, the device was encapsulated with a glass slide using UV-activated epoxy inside the glovebox.

**Spatially Resolved MEL**

The spatial imaging of MEL effects was captured via a scientific CMOS camera (Andor iStar) which was integrated into a microscopy setup. An infinity corrected objective (20x Mitutoyo Plan Apo) was used to magnify a micrometer scale region of the OLED for investigation. Measurements were performed under constant current (Keysight B2901A) and at room temperature. The OLED was suspended between the poles of an electromagnet (EM4-HVA) which was swept between positive and negative polarities with variable step sizes. Imaging was performed 3.5 seconds after each step to allow the field to stabilise. The intensity relative to the zero field, ΔEL/EL, was then calculated for the entire frame at each magnetic field value. Spatially resolved MEL was performed at each pixel in the CMOS sensor array, with a pitch between neighbouring points on the OLED of 0.30(5) μm. Fit parameters were then extracted at each position to generate the spatial parameter maps seen in Figure 3.

**Magnetic Resonance**

A Double Split-Square Resonator (DSSR) was designed to sit above the OLED while allowing light to pass through unimpeded to an optical system. This was achieved with a circular aperture positioned centrally to the DSSR's split squares as seen in Figure S2. Lock-in detection was utilised for both E/ODMR measurements. The OLED was operated under constant current conditions and at room temperature. A signal generator (SRS SG396) sent microwave pulses of width 10 μs and period 100 μs to the resonator after passing through a microwave amplifier. The strength of the resonance field at the OLED was ~ 0.01 mT.

In EDMR, the resulting periodic changes in device current were first amplified with a low-noise current amplifier (SRS SR570) utilising a 6 dB bandpass filter at 10 kHz before being detected by the lock-in amplifier (SRS SR865A). For ODMR, an optical fibre was fed through the resonator aperture to collect light emanating from the surface of the OLED. This was passed into a photodetector (Thorlabs PDA36A) with 60 dB gain. The output voltage was then further amplified (Femto DHPVA) with 50 dB gain before being passed into the lock-in detector.

**Spatial Autocorrelation – Moran's I**

Moran's I calculates an expected index value $\rho$ at each lag distance $r$ by taking the sum over all cross-products of elements in the set of spatial coordinates {i,j} (spatial autocovariance $\gamma_r$) and normalising it with respect to the spatial variance $\gamma_0$:

$$\rho_r = \frac{\gamma_r}{\gamma_0} = \frac{\sum_i^M \sum_j^N \sigma_{ij} \sum_k^M \sum_l^N w_{kl} \cdot \sigma_{kl}}{\sum_i^M \sum_j^N \sigma_{ij}^2}$$

where the standardised weight matrix $w_{kl}$ is a simple contiguity matrix defined as a two-dimensional lattice with column and row lengths reflecting the image data and is implicitly related to the lag distance $r_{ij}$. A correlogram is generated by calculating $\rho_r$ at each lag distance using the sensor lattice coordinates (i,j) and calibrating these to distances in the object plane (micrometers).

## Acknowledgements

We acknowledge support through the Australian Research Council Centre of Excellence in Exciton Science (CE170100026). We acknowledge the facilities and the scientific and technical assistance of Microscopy Australia at the Electron Microscope Unit (EMU) within the Mark Wainwright Analytical Centre (MWAC) at UNSW Sydney.

## Author Contributions

D.R.M. conceived the study. The experimental apparatus was developed by W.J.P., A.B. and A.M. Experiments were designed and performed W.J.P., and D.R.M. Devices were fabricated by R.G. and A.A. TEM and AFM were conducted by R.G. Data were analysed by W.J.P. and D.R.M., and all authors contributed to the discussion. W.J.P. and D.R.M. wrote the manuscript, with contributions from all authors.

## Data availability

The data that support the findings of this study are available within this paper and its Supplementary Information. Additional data are available from the corresponding author upon request.

## Competing Interest statement

The authors declare they have no competing interests.

# References

1. Warner, M. *et al.* Potential for spin-based information processing in a thin-film molecular semiconductor. *Nature* **503**, 504–508 (2013).

2. Baibich, M. N. *et al.* Giant magnetoresistance of (001)Fe/(001)Cr magnetic superlattices. *Phys. Rev. Lett.* **61**, 2472–2475 (1988).

3. Binasch, G., Grünberg, P., Saurenbach, F. & Zinn, W. Enhanced magnetoresistance in layered magnetic structures with antiferromagnetic interlayer exchange. *Phys. Rev. B* **39**, 4828–4830 (1989).

4. Dieny, B. *et al.* Giant magnetoresistive in soft ferromagnetic multilayers. *Phys. Rev. B* **43**, 1297–1300 (1991).

5. Orritt, M. & Brownt, R. Resonance in a Single Molecule. **363**, 244–245 (1993).

6. Lloyd, S. Universal Quantum Simulators. *Science (80-. ).* **273**, 1073–1078 (1996).

7. Maze, J. R. *et al.* Nanoscale magnetic sensing with an individual electronic spin in diamond. *Nature* **455**, 644–647 (2008).

8. Awschalom, D. D. & Flatté, M. E. Challenges for semiconductor spintronics. *Nat. Phys.* **3**, 153–159 (2007).

9. Gaita-Ariño, A., Luis, F., Hill, S. & Coronado, E. Molecular spins for quantum computation. *Nat. Chem.* **11**, 301–309 (2019).

10. Joshi, G. *et al.* Separating hyperfine from spin-orbit interactions in organic semiconductors by multi-octave magnetic resonance using coplanar waveguide microresonators. *Appl. Phys. Lett.* **109**, 103303 (2016).

11. Malissa, H. *et al.* Revealing weak spin-orbit coupling effects on charge carriers in a $\pi$-conjugated polymer. *Phys. Rev. B* **97**, 1–5 (2018).

12. McCamey, D. R. *et al.* Spin Rabi flopping in the photocurrent of a polymer light-emitting diode. *Nat. Mater.* **7**, 723–728 (2008).

13. Schott, S. *et al.* Polaron spin dynamics in high-mobility polymeric semiconductors. *Nat. Phys.* **15**, 814–822 (2019).

14. Zadrozny, J. M., Niklas, J., Poluektov, O. G. & Freedman, D. E. Millisecond coherence time in a tunable molecular electronic spin qubit. *ACS Cent. Sci.* **1**, 488–492 (2015).

15. Macià, F. *et al.* Organic magnetoelectroluminescence for room temperature transduction between magnetic and optical information. *Nat. Commun.* **5**, 1–7 (2014).

16. Kavand, M. *et al.* Discrimination between spin-dependent charge transport and spin-dependent recombination in π-conjugated polymers by correlated current and electroluminescence-detected magnetic resonance. *Phys. Rev. B* **94**, 075209 (2016).

17. Uoyama, H., Goushi, K., Shizu, K., Nomura, H. & Adachi, C. Highly efficient organic light-emitting diodes from delayed fluorescence. *Nature* **492**, 234–238 (2012).

18. Bayliss, S. L. *et al.* Optically addressable molecular spins for quantum information processing. *Science (80-. ).* **370**, 1309–1312 (2020).

19. Nguyen, T. D. *et al.* Isotope effect in spin response of π-conjugated polymer films and devices. *Nat. Mater.* **9**, 345–352 (2010).

20. Xiong, Z. H., Wu, D., Vardeny, Z. V. & Shi, J. Giant magnetoresistance in organic spin-valves. *Nature* **427**, 821–824 (2004).

21. Sun, D., Ehrenfreund, E. & Valy Vardeny, Z. The first decade of organic spintronics research. *Chem. Commun.* **50**, 1781 (2014).

22. Baker, W. J. *et al.* Robust absolute magnetometry with organic thin-film devices. *Nat. Commun.* **3**, 898 (2012).

23. Grünbaum, T. *et al.* OLEDs as models for bird magnetoception: Detecting electron spin resonance in geomagnetic fields. *Faraday Discuss.* **221**, 92–109 (2019).

24. Zhang, M., Höfle, S., Czolk, J., Mertens, A. & Colsmann, A. All-solution processed transparent organic light emitting diodes. *Nanoscale* **7**, 20009–20014 (2015).

25. Xu, R. P., Li, Y. Q. & Tang, J. X. Recent advances in flexible organic light-emitting diodes. *J. Mater. Chem. C* **4**, 9116–9142 (2016).

26. Bonizzoni, C. *et al.* Coherent coupling between Vanadyl Phthalocyanine spin ensemble and microwave photons: Towards integration of molecular spin qubits into quantum circuits. *Sci. Rep.* **7**, 1–8 (2017).

27. Joo, W. J. *et al.* Metasurface-driven OLED displays beyond 10,000 pixels per inch. *Science* **370**, 459–463 (2020).

28. Jamali, S., Joshi, G., Malissa, H., Lupton, J. M. & Boehme, C. Monolithic OLED-Microwire Devices for Ultrastrong Magnetic Resonant Excitation. *Nano Lett.* **17**, 4648–4653 (2017).

29. Malissa, H. *et al.* Room-temperature coupling between electrical current and nuclear spins in OLEDs. *Science (80-. ).* **345**, 1487–1490 (2014).

30. Frankevich, E. L. *et al.* Polaron-pair generation in poly(phenylene vinylenes). *Phys. Rev. B* **46**, 9320–9324 (1992).

31. Kalinowski, J., Cocchi, M., Virgili, D., Di Marco, P. & Fattori, V. Magnetic field effects on emission and current in Alq3-based electroluminescent diodes. *Chem. Phys. Lett.* **380**, 710–715 (2003).

32. Francis, T. L., Mermer, Ö., Veeraraghavan, G. & Wohlgenannt, M. Large magnetoresistance at room temperature in semiconducting polymer sandwich devices. *New J. Phys.* **6**, 1–8 (2004).

33. Sheng, Y. *et al.* Hyperfine interaction and magnetoresistance in organic semiconductors. *Phys. Rev. B* **74**, 045213 (2006).

34. Geng, R., Daugherty, T. T., Do, K., Luong, H. M. & Nguyen, T. D. A review on organic spintronic materials and devices: I. Magnetic field effect on organic light emitting diodes. *J. Sci. Adv. Mater. Devices* **1**, 128–140 (2016).

35. Boehme, C. & Malissa, H. Electrically detected magnetic resonance spectroscopy. *eMagRes* **6**, 83–100 (2017).

36. Mermer, Ö. *et al.* Large magnetoresistance in nonmagnetic π-conjugated semiconductor thin film devices. *Phys. Rev. B - Condens. Matter Mater. Phys.* **72**, 1–12 (2005).

37. Tetienne, J.-P. *et al.* Quantum imaging of current flow in graphene. *Sci. Adv.* **3**, e1602429 (2017).

38. Kim, D. *et al.* A CMOS-integrated quantum sensor based on nitrogen–vacancy centres. *Nat. Electron.* **2**, 284–289 (2019).

39. Hodges, M. P. P., Grell, M., Morley, N. A. & Allwood, D. A. Wide Field Magnetic Luminescence Imaging. *Adv. Funct. Mater.* **27**, 1606613 (2017).

40. Wang, F., Macià, F., Wohlgenannt, M., Kent, A. D. & Flatté, M. E. Magnetic Fringe-Field Control of Electronic Transport in an Organic Film. *Phys. Rev. X* **2**, 021013 (2012).

41. Segal, M., Baldo, A., Holmes, J., Forrest, R. & Soos, G. Excitonic singlet-triplet ratios in molecular

and polymeric organic materials. *Phys. Rev. B - Condens. Matter Mater. Phys.* **68**, 1–14 (2003).

42. Ehrenfreund, E. & Vardeny, Z. V. Effects of magnetic field on conductance and electroluminescence in organic devices. *Isr. J. Chem.* **52**, 552–562 (2012).

43. Kersten, S. P., Schellekens, A. J., Koopmans, B. & Bobbert, P. A. Magnetic-field dependence of the electroluminescence of organic light-emitting diodes: A competition between exciton formation and spin mixing. *Phys. Rev. Lett.* **106**, 1–4 (2011).

44. Schellekens, A. J., Wagemans, W., Kersten, S. P., Bobbert, P. A. & Koopmans, B. Microscopic modeling of magnetic-field effects on charge transport in organic semiconductors. *Phys. Rev. B - Condens. Matter Mater. Phys.* **84**, 1–12 (2011).

45. Fesser, K., Bishop, A. R. & Campbell, D. K. Optical absorption from polarons in a model of polyacetylene. *Phys. Rev. B* **27**, 4804–4825 (1983).

46. Baker, W. J., Keevers, T. L., Lupton, J. M., McCamey, D. R. & Boehme, C. Slow hopping and spin dephasing of coulombically bound polaron pairs in an organic semiconductor at room temperature. *Phys. Rev. Lett.* **108**, 1–5 (2012).

47. Crooker, S. A. *et al.* Spectrally resolved hyperfine interactions between polaron and nuclear spins in organic light emitting diodes: Magneto-electroluminescence studies. *Appl. Phys. Lett.* **105**, 153304 (2014).

48. Sheng, Y. *et al.* Hyperfine interaction and magnetoresistance in organic semiconductors. *Phys. Rev. B - Condens. Matter Mater. Phys.* **74**, (2006).

49. McCamey, D. R. *et al.* Hyperfine-Field-Mediated Spin Beating in Electrostatically Bound Charge Carrier Pairs. *Phys. Rev. Lett.* **104**, 017601 (2010).

50. Bayliss, S. L., Greenham, N. C., Friend, R. H., Bouchiat, H. & Chepelianskii, A. D. Spin-dependent recombination probed through the dielectric polarizability. *Nat. Commun.* **6**, 8534 (2015).

51. Bergeson, J. D., Prigodin, V. N., Lincoln, D. M. & Epstein, A. J. Inversion of magnetoresistance in organic semiconductors. *Phys. Rev. Lett.* **100**, 1–4 (2008).

52. Wang, Y., Sahin-Tiras, K., Harmon, N. J., Wohlgenannt, M. & Flatté, M. E. Immense magnetic response of exciplex light emission due to correlated spin-charge dynamics. *Phys. Rev. X* **6**, 1–12 (2016).

53. Waters, D. P. *et al.* The spin-Dicke effect in OLED magnetoresistance. *Nat. Phys.* **11**, 910–914 (2015).

54. Kanemoto, K., Hatanaka, S. & Suzuki, T. Correlation between bias-dependent ESR signals and magnetic field effects in organic light emitting diodes. *J. Appl. Phys.* **125**, 125501 (2019).

55. Bayat, K., Choy, J., Farrokh Baroughi, M., Meesala, S. & Loncar, M. Efficient, uniform, and large area microwave magnetic coupling to NV centers in diamond using double split-ring resonators. *Nano Lett.* **14**, 1208–1213 (2014).

56. Xie, G. *et al.* Measuring and structuring the spatial coherence length of organic light-emitting diodes. *Laser Photon. Rev.* **10**, 82–90 (2016).

57. Geng, R., Pham, M. T., Luong, H. M., Short, A. & Nguyen, T. D. Correlation between the width and the magnitude of magnetoconductance response in π-conjugated polymer-based diodes. *J. Photonics Energy* **8**, 1 (2018).

# Spatial Variation and Correlation of Spin Properties in Organic Light-Emitting Diodes – Supporting Information

W. J. Pappas[1], R. Geng[1], A. Mena[1], A. Baldacchino[1], A. Asadpoordarvish[1], D. R. McCamey[1,*]

[1]ARC Centre of Excellence in Exciton Science, School of Physics, UNSW Sydney, NSW 2052, Australia

*Corresponding author:  dane.mccamey@unsw.edu.au

# 1  Experimental Setup

A novel approach to the detection of magnetoelectroluminescence (MEL) was implemented by incorporating optical microscopy into a traditional MEL setup in order to spatially resolve the intra-device variations of this effect. This was accomplished by suspending an organic light emitting diode (OLED) in a homogeneous magnetic field and below an optical microscope (Mitutoyo M Plan APO 20x) which was connected to a scientific CMOS camera (Andor iStar sCMOS 18UA3) – see Figure S1a and S1b. A sample stage was constructed to house the OLED in conjunction with a microwave resonator to enable additional magnetic resonance measurements to be undertaken, while allowing emitted light to pass through to the objective unimpeded (Figure S1c and S1d). The optical path of the parallel light beam was then deflected by a silver-coated mirror angled at 45 degrees and was subsequently refocused using a tube lens onto the camera's photocathode.

The sample stage houses three main components – the OLED, a printed circuit board to power the device, and finally the resonator. Figure S1c shows how each of these were incorporated to achieve the design goals of the setup to enable spatially resolved MEL. The structural components of the stage were 3D printed using PLA filament, except for the four non-magnetic brass rods. The power board and resonator were both fabricated using copper striplines deposited onto FR4 dielectric material. The manual translation stage (seen in the first 3 sub-panels of Figure S1c) was later replaced with a piezoelectric 2D translation stage (Micronix PPX-32CR) which enabled transverse movements of the OLED with high fidelity (10 nm), allowing us to target specific intra-device regions with high precision (sub-panel iv).

Microscopic imaging of the OLED electroluminescence (EL) at several current densities was captured on the sCMOS camera for numerous regions of the device. To ensure consistency between datasets examined in this investigation, we fixed the region of interest of the device while adjusting the exposure time of the camera for each of the OLED current densities to keep the detected EL (number of counts in the electronic well depth) of the sCMOS sensor roughly constant ($\sim 4.5 \times 10^4$ counts). We note that measurements were taken in a light-insulated environment and without the need for further amplification by the image intensifier of the camera.

The design of our Double Split-Square Resonator (DSSR) was adapted from a similar split-ring resonator fabricated by Bayat et al.[1]. The layout of our resonator produced a uniformly oscillating magnetic field, exhibiting efficient coupling to our spin system over a millimetre-squared area. This ensures that a large ensemble of spins is excited over the sample volume of the OLED. Moreover, we generate a low-GHz spectrum microwave field to avoid impacts from spin-orbit coupling to our magnetic resonance spectrums. Employing FR4 as the dielectric with a thickness of 1.6 mm, the first mode exhibited a resonance dip at 2.2855 GHz and is a regime capable of investigating the direct impacts of hyperfine couplings between charge carrier and nuclear spins in OLEDs. A microwave field strength of $B_1 \sim 0.01$ mT was measured using a calibrated Beehive Electronics (100B) probe, confirming that magnetic resonance was probed in the weak perturbative limit ($B_1 < |\Delta B_{hf}|$). We note that the resonator was operated under a quasi-continuous wave regime whereby the microwave field was pulsed with a period of 100 μs and a 10% duty cycle to minimise dissipative heat losses.

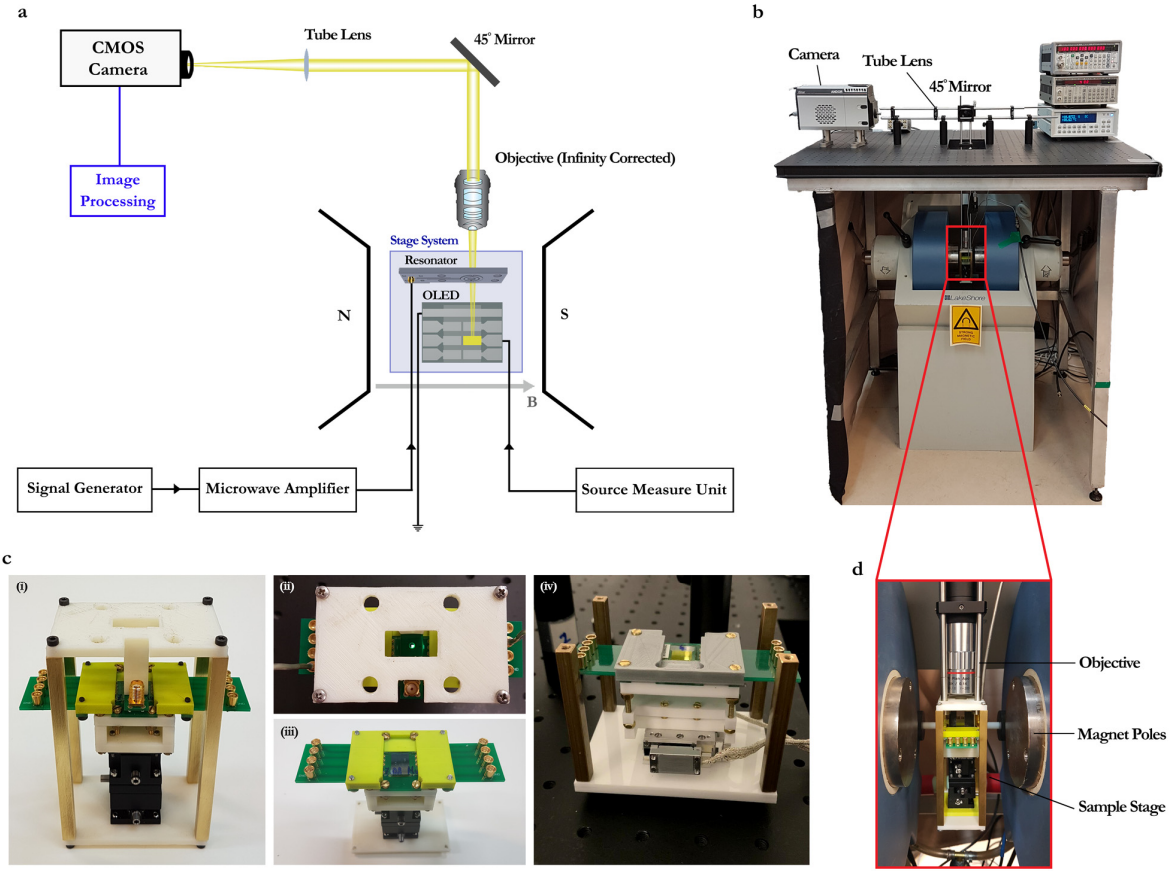

**Figure S1**: The measurement system. a) Schematic of the experimental setup detailing how the microscopy system was integrated into the MEL setup. Light passes through the objective and is directed and refocused onto the sensor of the sCMOS camera. b) Photograph of the experimental setup. The blow up (d) shows the OLED housed in the sample stage between the two electromagnet pole pieces and suspended centrally below the objective. (c) Photographs of sample stage. (i) A front view of the sample stage. Starting from the bottom, we have a linear 2D-translation stage, attached to this is a power board which allows electrical current to pass through to the OLED, the OLED, and finally held overhead is the double split-square resonator which is mounted ~1 mm above the OLED. (ii) A top view showing light emanating from a test OLED. (iii) The sample stage with the top panel and resonator removed. (iv) A motorised 2D-piezoelectric stage replaced the manual translation stage and allowed for transverse movements of the OLED under the imaging system.

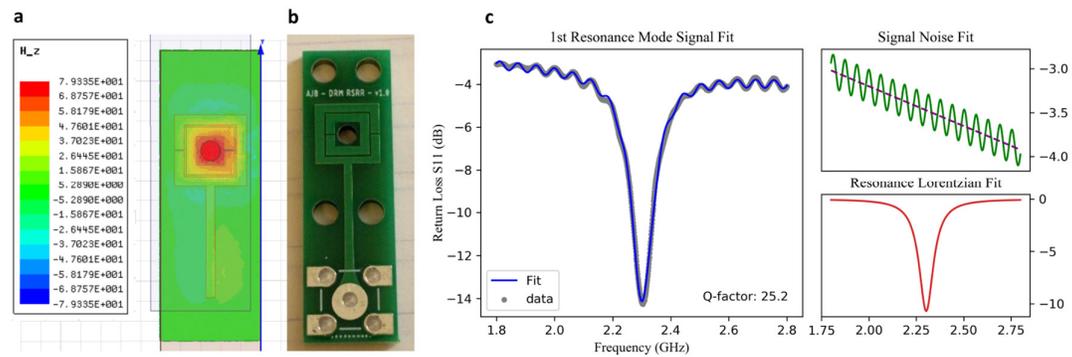

**Figure S2**: a) Simulated microwave fields generated by the modelled DSSR exhibiting high uniformity around the central aperture. b) Fabricated DSSR containing the four mounting holes, vias for SMP connection, and mounting pads

for easy soldering. c) Decomposition of the return loss signal (S11) from the first resonant mode of the DSSR. The raw signal captured by the network analyser, decomposed into its background component (upper right) and resonance component (lower right). The Q-factor, which is calculated from the Lorentzian fit, is 25.2, and the first resonant mode is centred on a frequency of ~ 2.3 GHz.

## 2 OLED Characterisation

Thin-film homogeneity of the active layer is a crucial component in understanding the intra-device variation of spin-properties in the device. Figure S3a shows an image taken with an optical microscope of the OLED used in this study undergoing EL at a current density of 1.67 mA/cm$^2$ and exhibiting large area uniformity in emission. This reflects the excellent film uniformity of the organic (SY-PPV) layer which is defect-free over large areas. The spectrum of this device shows a single emission band with a peak wavelength ~600 nm (Figure S3b). The device J-EL-V characteristics were measured over the range of device currents used in the investigation and show typical diode behaviour (Figure S3c).

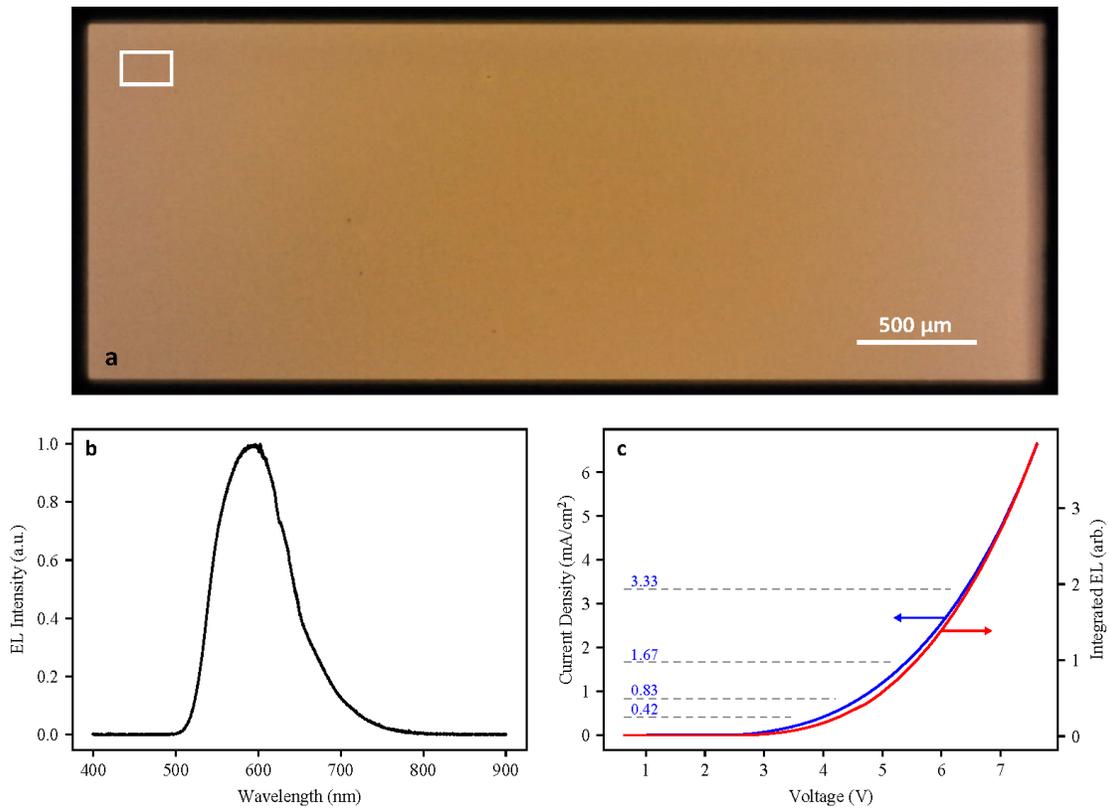

**Figure S3**: Characterization of the SY-PPV OLED. (a) An image of the OLED undergoing electroluminescence at a current density of 1.67 mA/cm$^2$ using an optical microscope which shows excellent film uniformity. The area of the device is 1.5 mm x 4 mm. The white box shows the region (190.8 x 122.1 µm) over which MEL was resolved. (b) EL spectrum showing single-band emission with a peak wavelength of 596.5 nm. (c) J-EL-V characteristics. The horizontal dashed lines cutting the J-EL-V curves show the four current density measurements (doubling with each increment) used in the spatially resolved analysis.

**Film Morphology**

We now investigate the film homogeneity of SY-PPV on a much smaller scale in both the vertical and transverse directions to consider variations in the film thickness and morphology as the source of the ≳ 7 μm correlation structures observed in the spatial parameter maps. Importantly however, MEL is an effect which originates predominantly in the bulk of the active (SY-PPV) layer, and therefore any morphological surface studies are of limited efficacy in explaining the origin of the correlated magnetic field effect behaviour.

We utilise transmission electron microscopy (TEM) to study variations in the film thicknesses for each of the layers deposited sequentially in the device structure. The SY-PPV film, where the magnetic field effects of our spatial maps occur, shows excellent uniformity at the metal-organic interface and is homogeneous throughout the layer (Figure S4). Note that discerning the interface between the PEDOT:PSS and SY-PPV films is ambiguous with TEM – a technique sensitive to the atomic weight of the target material – as both organic materials are comprised of similarly lightweight atoms. Moreover, as the thicknesses of deposited layers are small relative to observed spatial autocorrelations, we instead turn our attention to variations in the transverse plane where the surface area greatly exceeds the measured correlation lengths.

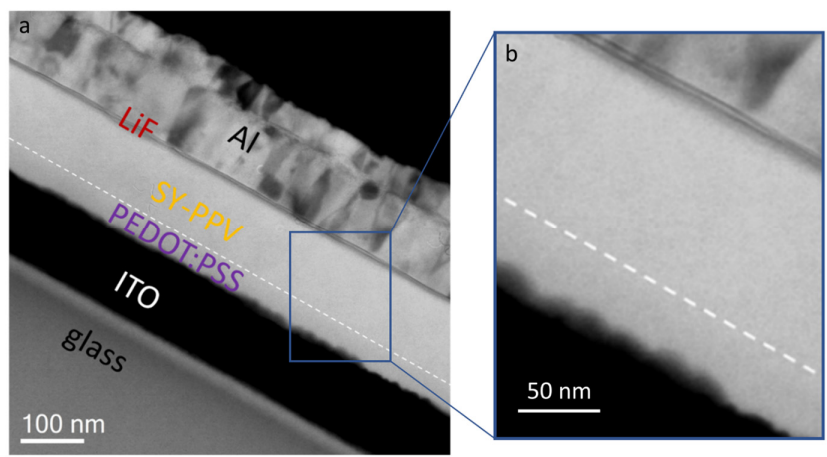

**Figure S4**: a) TEM image of the cross-section of the SY-PPV OLED: glass/ITO (~115nm)/PEDOT:PSS (~ 32nm)/SY-PPV (~110nm)/LiF (~1nm)/Al (~150nm). We note that the white dash line indicates the approximate location of the interface between the PEDOT:PSS and SY-PPV layer, which could not be clearly distinguished under TEM. The thicknesses of the PEDOT:PSS and SY-PPV layers were characterized separately.

In order to study the surface morphology of the thin film comprising the active layer, we employ atomic force microscopy (AFM) on SY-PPV films both with (Figure S5a) and without (Figure S5b) the buffering PEDOT:PSS layer. When SY-PPV is directly deposited onto the ITO, no clear spatial features are observed, and the surface morphology appears uniform and random. However, some spatial structure is introduced at the surface upon the inclusion of a thin PEDOT:PSS layer deposited below the SY-PPV layer and is likely the result of aggregation in the PEDOT:PSS film. Spatial autocorrelation analysis performed on the AFM datasets yield statistically significant correlation lengths of below 150 nm (ITO/SY-PPV) and ~ 5 μm (ITO / PEDOT:PSS / SY-PPV) as displayed in Figure S5c and d, respectively. Despite the latter exhibiting a correlation length similar to those of the MEL parameters, the lineshape of the AFM correlogram is substantially different, showing a much more rapid decay than those of A and $B_{1/2}$ (Figure S7). While further investigation into the effects of the PEDOT:PSS layer on the surface morphology of the active layer may be warranted, the characteristically different correlogram, along with the observation of these magnetically-induced spatial correlations in similar small-molecule OLEDs excluding PEDOT:PSS from the device architecture entirely (Figure S8), indicates that the surface morphology is not the primary source of spatially correlated magnetic field effects in OLED devices.

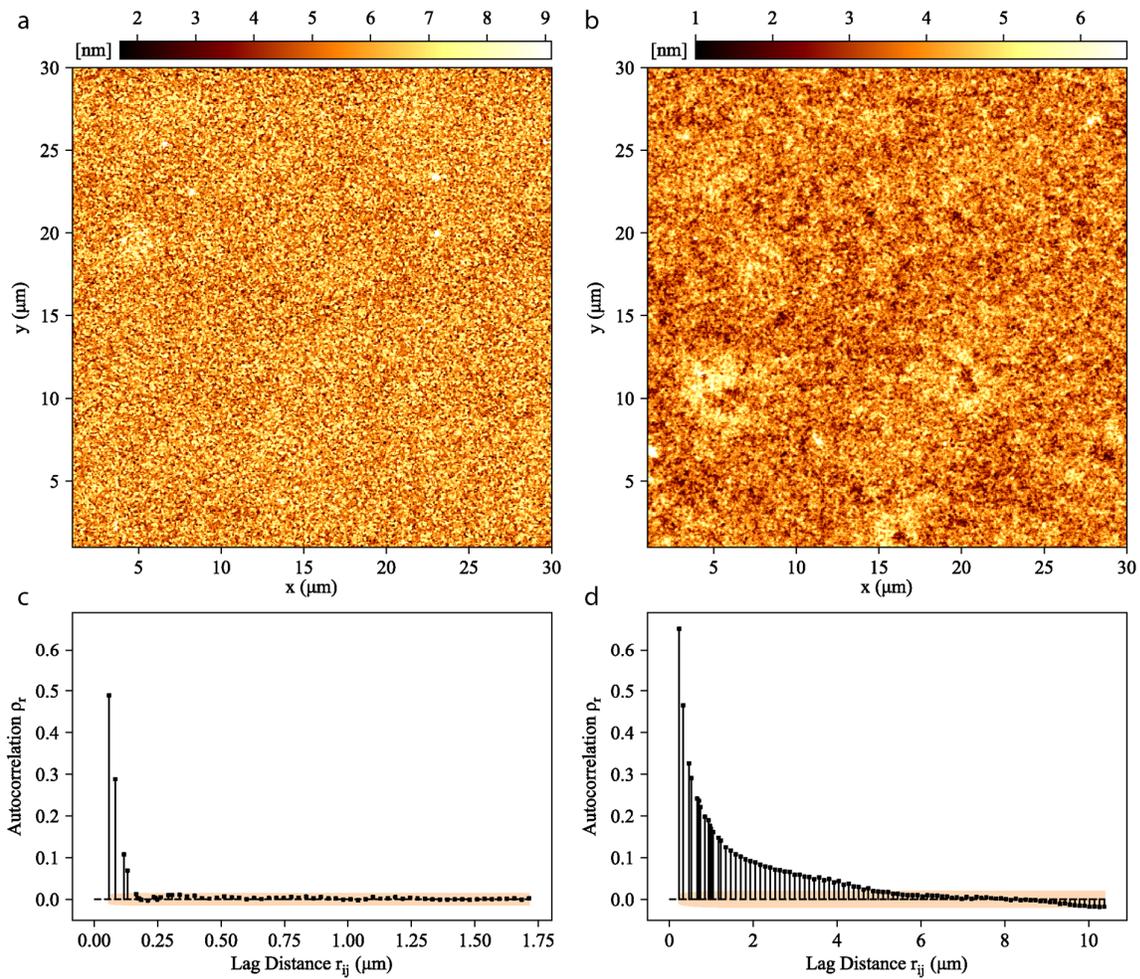

**Figure S5:** Surface morphology characterization of (a) ITO/SY-PPV and (b) ITO/PEDOT:PSS/SY-PPV films using atomic force microscopy (AFM). (a) The surface of the SY-PPV film exhibits a uniform and random morphology. (b) Spatial structure is observed in the surface of the SY-PPV film when a thin PEDOT:PSS layer is deposited below, with aggregation of the PEDOT:PSS material likely explaining this behaviour. Spatial autocorrelation analysis is carried out on each of the AFM maps, (c) and (d) respectively. A small spatial correlation on the ITO/SY-PPV surface is observed which rapidly decays at distances < 150 nm, while a substantial correlation length of ~5 μm is measured on the ITO/PEDOT:PSS/SY-PPV surface.

## Emission Spatial Autocorrelation

In addition to the surface morphology examinations carried out on inactive devices, we have investigated the homogeneity of electroluminescent emission of the SY-PPV OLED while under operation. Here, we consider whether the spatially resolved zero-field emission is related to the magnetic field induced change in intensity, while also demonstrating that the spatially correlated spin properties display behaviour different to that arising from the surface morphology of the layer where these effects occur.

## Intensity Maps

Maps of the OLED intensity over a range of magnetic fields can be studied to extract useful information regarding the emission properties of the device, while allowing us to generate the ΔEL/EL maps necessary to perform spatially resolved MEL analysis.

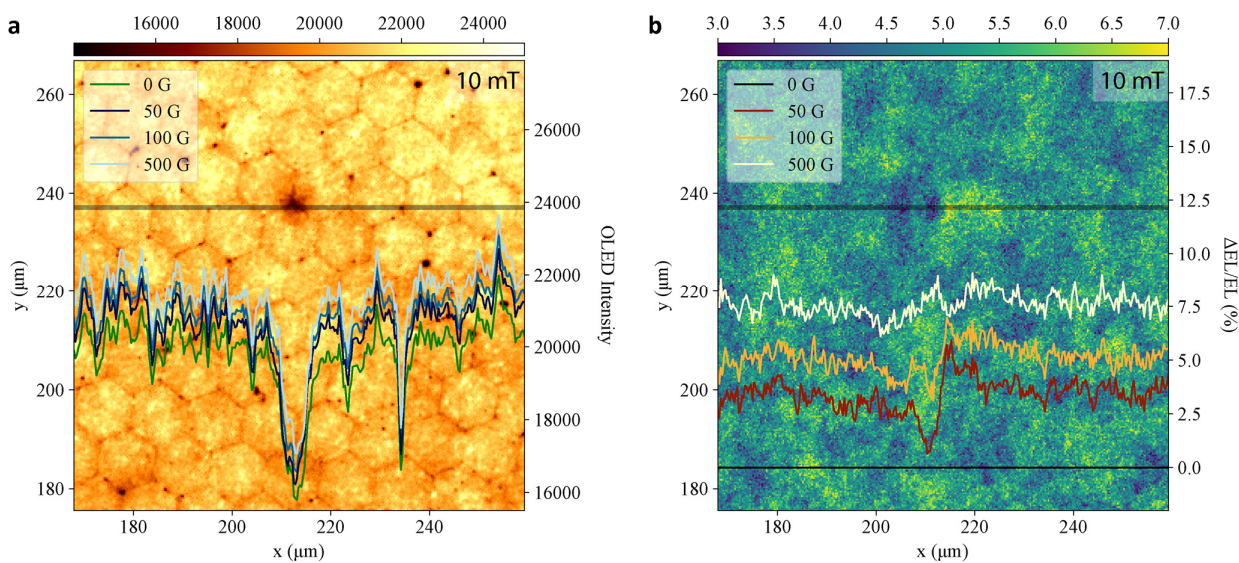

**Figure S6:** Emission intensity maps with intensity slices for several magnetic field strengths. (a) A spatially resolved OLED electroluminescence map at 10 mT, with the intensity slices depicting the emission along the (horizontal) x-axis for the area encompassed within the dark band (thickness of 1 μm) and for several magnetic field values. The hexagonal pattern arises due to the bundling of fibres and is a fixed feature of the camera's sensor. Also, the small dark spots present at some of the vertices of these bundles are trapped dust particles which lie stationary on the sensor. These are distinguished from impacts to the OLED film homogeneity (resulting in low-EL emission) which are characterised by a wide area intensity drop-off, as seen at roughly 210 μm along the horizontal intensity band. (b) The ΔEL/EL map spatially resolved over the same region at 10 mT. Importantly, any artefacts introduced by the camera, which remain unchanged with time and magnetic field strength, are subsequently subtracted out of the map when calculating the intensity change relative to the zero-field frame. These maps therefore provide an excellent insight into the intra-device variation of magnetic field dependent EL. The ΔEL/EL intensity slices for the same set of field values in (a) are displayed. Interestingly, the defect located at 210 μm shows complicated behaviour that differs from the more homogenous sections of the OLED.

The hexagonal structure seen in the OLED intensity map (Figure S6a) corresponds to the edges of bundled fibres that are tightly packed to form the fibre optic coupler which serves to couple the image intensifier to the CMOS sensor. The separation gap between fibre bundles (< 1 μm in the object plane) is a fixed feature of fibre optic plate and is therefore entirely absent from the ΔEL/EL maps, meaning that it does not impact our spatially resolved MEL data (see also Figure S7). Moreover, the small dark spot-like features present at several vertices between fibre bundles are due to the presence of dust lying on the fibre optic plate and are well defined. By comparison, impacts to the film homogeneity resulting in areas of low EL are distinguished by their shape and wide area intensity drop-off, as observed at roughly 210 μm along the horizontal intensity band. We can use impacted areas of the film like these to definitively show that no substantial magnetic-dependent movement of the OLED takes place over the course of the measurement (besides isotropic jitter < 1 μm likely resulting from vibrations). Accurate tracking of the OLED is accomplished by comparing the intensity slices at various field strengths around these regions which exhibit large intensity gradients and are therefore highly sensitive to positional changes.

We also perform spatial autocorrelation analysis on the zero-field OLED intensity ($I_{B=0}$) map (over the larger region used for MEL analysis in the main text) and compare the resulting correlogram to those calculated from the MEL parameter maps. A clear distinction in the lineshapes of morphology-related correlograms (AFM, $I_{B=0}$) is observed compared to those characterising the magnetic field effect parameters. We use this to conclude that the surface morphology and the zero-field intensity are not likely to be the origin of the spatially correlated magnetic field effects.

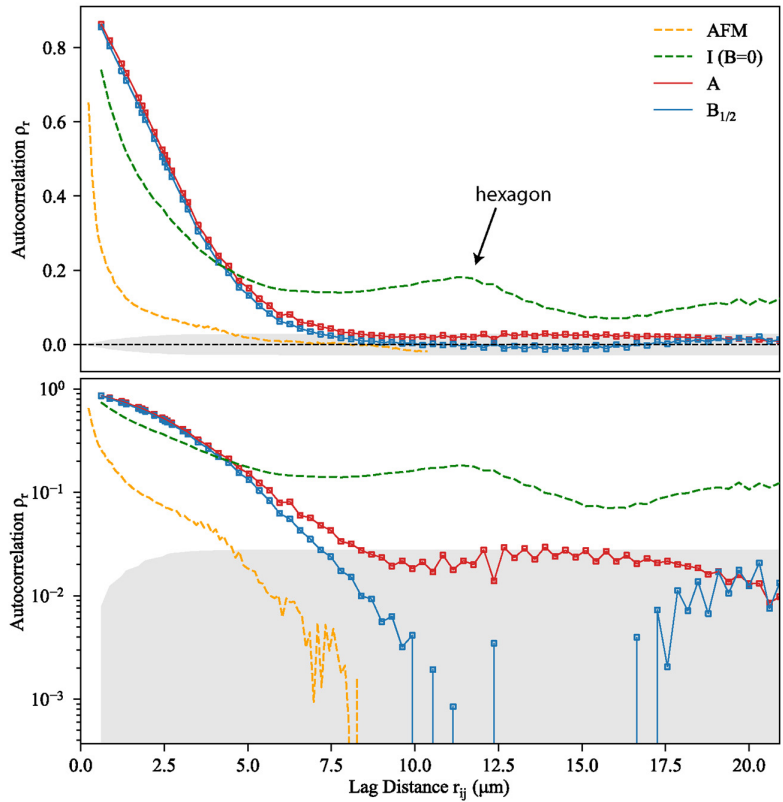

**Figure S7**: A comparison of the spatial autocorrelation of the ITO/PEDOT:PSS/SY-PPV surface morphology (AFM), the zero magnetic field emission intensity ($I_{B=0}$), and the MEL spatial maps for the amplitude (A) and half-width ($B_{1/2}$) parameters. The same correlograms are plotted on a log scale (lower panel) which better show the rate of decay for each of the spatial autocorrelations. A clear similarity between lineshapes of the MEL parameter correlograms is observed, whereas the AFM and zero-field intensity correlograms exhibit substantially different lineshapes and correlation lengths. From this comparison, we conclude that the surface morphology and the zero-field EL intensity of the device are likely not the primary sources of the spatial structure observed in the magnetic field effect maps. We also note that the experimental artefact pertaining to the fibre optic coupler of the camera (hexagonal boundaries), which is present in the zero-field EL intensity map and manifests itself as a positive bump in autocorrelation at 12 μm and 24 μm in the $I_{B=0}$ correlogram, is a feature which is entirely absent in the A and $B_{1/2}$ correlograms.

Lastly, we compare the spatial autocorrelation of the $B_{1/2}$ magnetic field parameter of an Alq$_3$ OLED under both 5x and 20x optical magnification. This allows us to unambiguously rule out any effect from the hexagonal structure of the fibre optic coupler (which would occur at a fixed pixel separation distance in the CMOS sensor under both magnifications), while allowing us to independently calibrate length scales in the object plane with high fidelity. The correlograms in Figure S8 demonstrate a spatial autocorrelation that strongly depends on the physical distance in the object plane (bottom x-axis) rather than on the pixel separation distance (top x-axes). Images captured under the higher 20x magnification also have the advantage of resolving the correlated MEL structures at smaller distances and with higher precision. However, we note that correlation values which lie below the optical system's limit of resolution of roughly 1.3 μm are likely to be inflated. Nonetheless, our correlation lengths for both A and $B_{1/2}$ lie well above this threshold.

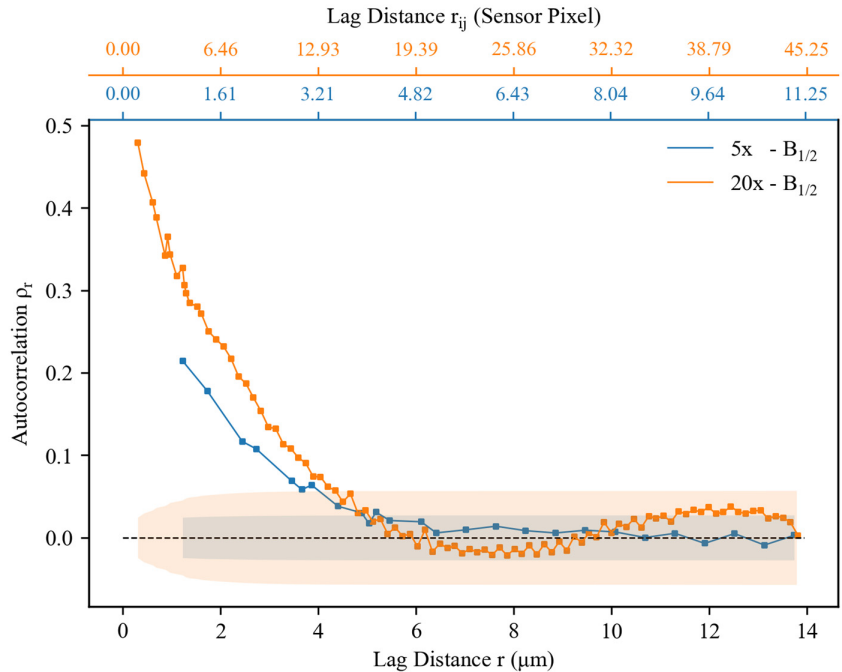

**Figure S8:** Spatial autocorrelation of a test (Alq$_3$) OLED under 5x (blue) and 20x (orange) optical magnification. Reassuringly, the separately calibrated length scales for each magnification produce correlograms which exhibit a similar statistically significant correlation length of roughly 5-6 μm. The corresponding lag distances calculated in terms of neighbouring sensor pixel ($r_{ij}$) are displayed on the top axes. Note that $r_{ij}$ distances (in pixel number) in the correlogram were binned in 0.5 intervals at larger distances, explaining why the 20x scale contains 4 times the pixel count plus an additional half of the bin size.

In addition to excluding features in the detection system, we removed the PEDOT:PSS buffering layer (in place of a thicker hole transport layer) from the Alq$_3$ OLED measured under 20x magnification to exclude this as the source of our correlated MEL structures. We point to the clear agreement in statistically significant correlation lengths (~ 5 μm) between devices both with and without PEDOT:PSS to verify that this effect is independent of the buffering layer material. This provides further supporting evidence to the conclusion drawn from Figure S7 – namely, that the spatially correlated magnetic field effects do not originate from the surface morphology imposed on the SY-PPV layer from underlying layers as measured in the AFM data (Figure S5b).

# 3 MEL Fitting

We employ the specific Non-Lorentzian (NL) two-parameter model to fit our MEL curves as outlined in the main text. The external magnetic field is swept over both polarities during these measurements, and while monolithic MEL produce highly symmetrical lineshapes around the zero-field (Figure 1d) with almost identical amplitudes and half-widths (< 2% change), microscopic MEL curves instead exhibit substantial local variations (> 40%) between positive and negative field components over the same regions in space. Fit qualities remain excellent over these highly resolved areas ($R^2 \sim 0.97$ – see Figure S9d), with the spatial groupings of similarly varying lineshapes (see clusters in parameter maps – Figure S14) indicating that parameters undergo a temporal evolution between measurements. This time dependence is discussed in further detail in Section 4, and we conclude by performing fits to MEL curves belonging only to one polarity of the field sweep in order to minimise any dynamic effects occurring in the parameter spaces and therefore improve mapping accuracy.

**Spatially Resolved MEL Fits**

As the spatially correlated MEL structures are observed at small length scales, a sufficiently high signal-to-noise ratio becomes critical to achieve accurate MEL fitting. As observed in the histogram current series of Figure 3c and 3d, the lower fit quality at higher current densities (as measured by the residuals in the $R^2$ 'goodness of fit' method) leads to some broadening in the parameter distributions (Table S1). This is particularly apparent with the growth of a heavy-tailed or positive skew in the $B_{1/2}$ distribution at 1.67 and 3.33 mA/cm$^2$, where we believe that electrical straining of the organic material at higher voltages plays a role in its emissive stability. Using spatially resolved MEL data from the lowest current density of 0.42 mA/cm$^2$, we attain high quality fits to each of the single pixel MEL curves (Figure S9d). The coefficient of variation (CV) is also calculated (Figure S9a) and any fits resulting in a $B_{1/2}$ parameter uncertainty exceeding 10% are ignored, with the $B_{1/2}$ spatial map plotted excluding these points (white spaces) in Figure S9b. Correlated structures remain largely unaffected, verifying that these regions do not arise from fitting uncertainty. Moreover, the normalised distributions both with and without points exceeding a CV of 10% are displayed in Figure S9c, where no change in the $B_{1/2}$ distribution takes place.

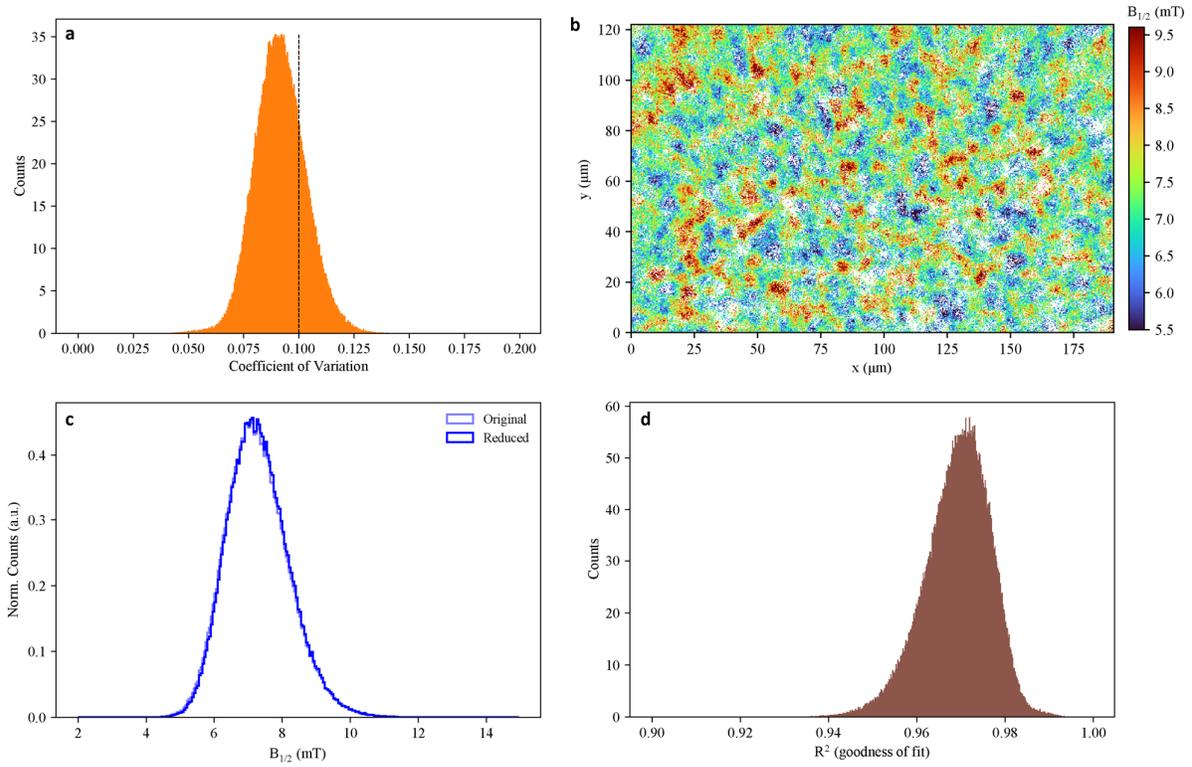

**Figure S9**: Quantifying the fit uncertainties for all single pixel MEL fits in the 0.42 mA/cm$^2$ dataset displayed in Figure 3 of the main text. (a) The coefficient of variation (CV) histogram displaying the percentage uncertainty of the $B_{1/2}$ parameter in the MEL fit. A threshold of 10% is chosen to investigate the effect of noise on the dataset. (b) A reduced $B_{1/2}$ spatial map with points removed (replaced with white spaces) that have a CV exceeding 10%. Such points are located mainly in the intermediate zones between small and high valued $B_{1/2}$ parameters. This verifies that the correlated structures are not an artefact arising from fits with higher uncertainties. (c) The normalised $B_{1/2}$ distributions for both original and reduced spatial maps. The distributions remain unchanged, again showing that the clustering of extreme valued parameters is not a result of fitting uncertainty. (d) The $R^2$ 'goodness of fit' histogram quantifying the quality of fits to the single pixel MEL curves. High quality fits are demonstrated which are distributed around a peak $R^2 = 0.97(3)$.

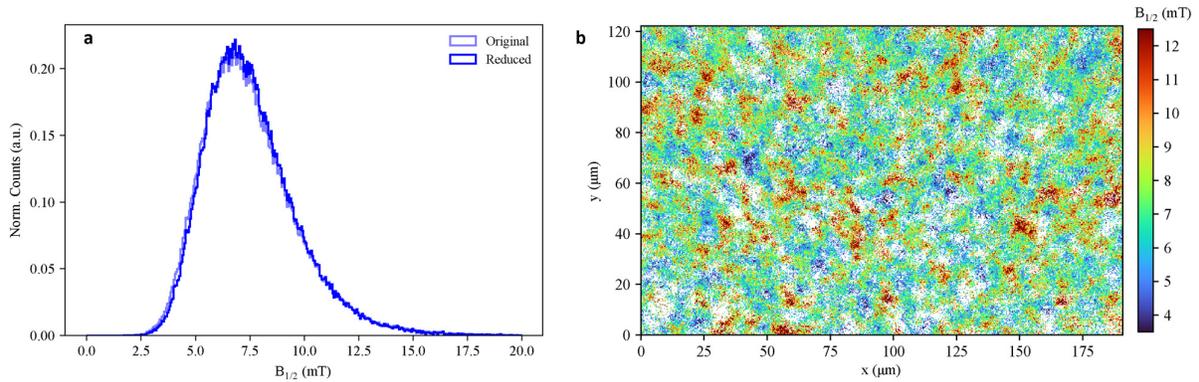

**Figure S10**: The reduced $B_{1/2}$ (a) distribution and (b) spatial map for the highest current density of 3.33 mA/cm$^2$. The fit quality is maximally distributed around $R^2 \sim 0.8$ and the $B_{1/2}$ CV histogram at $\sim 0.21$. We define the CV threshold at 0.25 and include fits with values only below this. When comparing the reduced $B_{1/2}$ distribution to the entire distribution, the heavy-tailed lineshape remains unaffected, indicating that its growth at higher operating biases is unrelated to the poorer quality fits in the distribution.

**MEL Parameter Distributions**

When measuring the intra-device variability of each MEL parameter (A, $B_{1/2}$), we employ a method of relative variability which measures the size of the full-width at half-maximum (FWHM) in relation to its mean – i.e. $A_{FWHM} / A_{mean}$. This method is well suited to describing the variability of $B_{1/2}$ distributions whose means remain unchanged by current density. Table S1 contains the relevant fitting parameters of the A and $B_{1/2}$ histograms fit to normal distributions. Included are the FWHM values, which are useful as an absolute measure of variability (particularly for the amplitude parameter), as well as bulk fits to monolithic measures of MEL for comparison. We also point out that the two highest current density $B_{1/2}$ histograms exhibit heavy-tailed lineshapes (see Figure 3d) which grow with increasing current density.

Table S1: MEL parameter distribution fits.

|  | 0.42 mA/cm² | 0.83 mA/cm² | 1.67 mA/cm² | 3.33 mA/cm² |
|---|---|---|---|---|
| **Amplitude (A)** | [%] | | | |
| Bulk Parameter | 8.98 | 7.44 | 6.05 | 4.78 |
| Mean | 9.00 | 7.46 | 6.09 | 4.84 |
| FWHM | 1.00 | 1.04 | 1.10 | 1.14 |
| $A_{FWHM}/A_{Mean}$ [%] | 11.2 | 13.9 | 18.0 | 23.5 |
| **Half-width ($B_{1/2}$)** | [mT] | | | |
| Bulk Parameter | 7.21 | 7.22 | 7.18 | 7.19 |
| Mean | 7.18 | 7.15 | 7.08 | 7.04 |
| FWHM | 2.10 | 2.56 | 3.49* | 4.44* |
| $B_{1/2,FWHM}/B_{1/2,Mean}$ [%] | 29.2 | 35.8 | 49.3* | 63.0* |

*Note that measurements denoted by an asterisk are fit with a Gaussian profile however exhibit heavy-tailed lineshapes.

Spatially averaging ΔEL/EL data allows us to improve our signal-to-noise ratio of the MEL curves and obtain slightly better fits. We can achieve this by binning the detected electroluminescence of neighbouring N x N pixels (Bin N). One consequence of binning is that spatial resolution is reduced resulting in the loss of spatial information, and hence a narrowing of the $B_{1/2}$ distribution. Therefore, there is a trade-off between losing spatial information by binning (averaging) the ΔEL/EL data and reducing the impact of fitting uncertainty of the MEL curves which instead broadens the distribution.

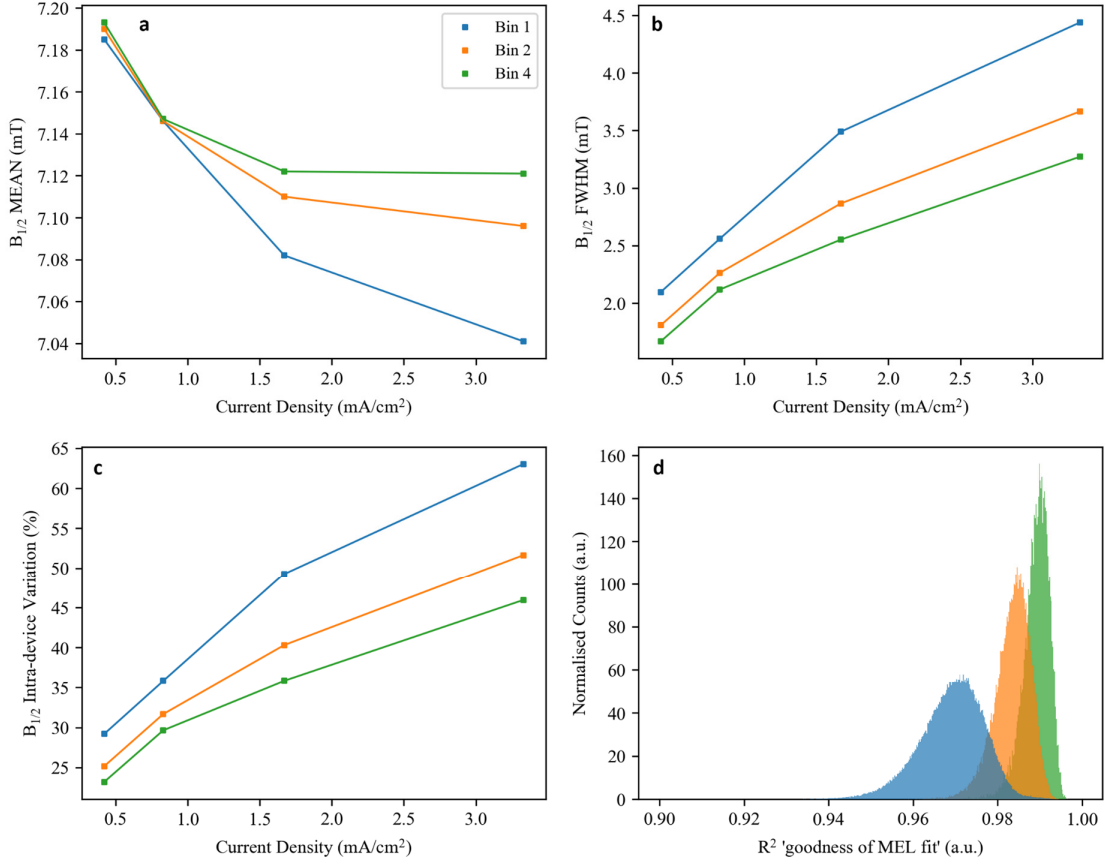

**Figure S11**: Spatial binning and device current density trends of the $B_{1/2}$ distribution. Spatial bins were performed by averaging neighbouring N x N pixel regions (Bin N) of the ΔEL/EL frames at each magnetic field strength. The (a) mean and (b) FWHM of the $B_{1/2}$ distribution for four current densities and three sets of spatially binned ΔEL/EL data. (c) Intra-device variation of the $B_{1/2}$ parameter exceeding 20% and is primarily influenced by a broadening of the distribution (FWHM) with the mean remaining relatively unchanged. We note that the minimum variability in $B_{1/2}$ observed at the lowest current density remains large for the very high confidence fits at Bin 4 averaging despite the loss of spatial information leading to a narrowing in the distribution. (d) Fits to the binned MEL curves are made, with larger bins improving the signal-to-noise ratio and yielding overall better fits.

## 4 Parameter Maps and Correlations

Light emission from OLEDs is a complicated process whose local variation in intensity shows a time-dependence which is likely sensitive to the molecular and nuclear spin reorganisation times[2,3]. In order to get a handle on the dynamics of the processes leading to changes in the magnetically induced EL, we track the individual ΔEL/EL frames which comprise a single spatially resolved MEL dataset. Figure S12 displays the 'deviation from the mean' of ΔEL/EL maps for a limited range of magnetic field strengths taken sequentially (with 27 seconds elapsing between successive frames). From a qualitative point of view, similar structures can be tracked for neighbouring δ(ΔEL/EL) maps which slowly evolve into different structures over several maps. This can be quantified by calculating the Pearson correlation coefficient between intensities of a selected map (at 7.0 mT) to those at the same positions in all other maps in the field sweep.

The correlation between spatially resolved ΔEL/EL and absolute OLED intensities compared to those at 7.0 mT are displayed in Figure S13a and b, respectively. As the magnetic field step size increases at larger fields, we also display correlation coefficients as a function of measurement time over the field sweep in Figure S13c and d. Using the larger correlations exhibited at frames around 7.0 mT, we estimate that the processes involved in emission for both magnetic and non-magnetic induced EL remain similar for timescales of at least minutes. This provides some insight as to why correlated structures in parameter maps from successive MEL measurements do not produce clustering in similar regions despite yielding almost identical distributions. We note however that temporally resolved correlations are complicated as they depend on both the measurement time and external magnetic field, and a thorough investigation into de-coupling this dependence is not undertaken here.

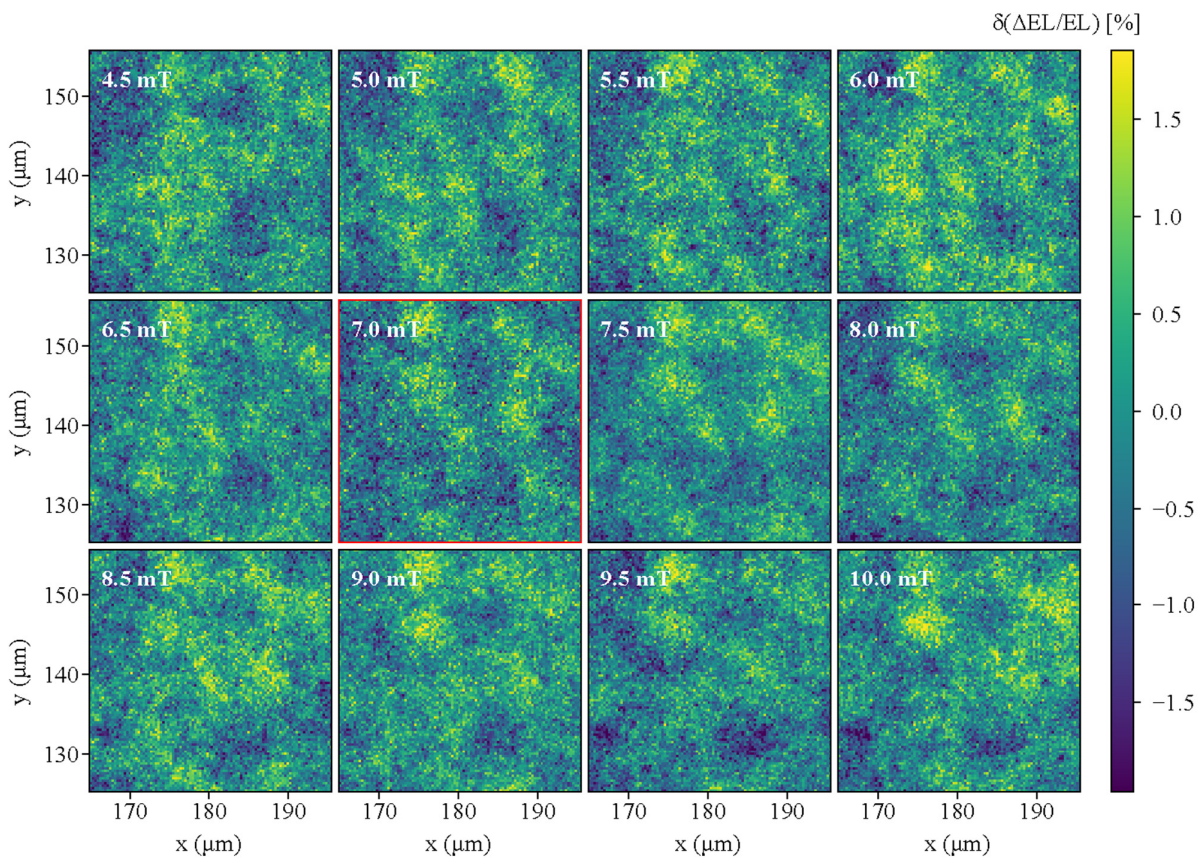

**Figure S12**: A sequence of 'deviation from the mean' ΔEL/EL maps for successive measurements over a limited range of the magnetic field sweep. Only deviations from the mean ΔEL/EL value for each measurement are of interest as they result in the correlated structures observed in the MEL parameter maps after each point is fit. Similar structure between neighbouring frames can be seen which evolves over several field values that occurs on the timescale of several minutes. We quantify this change in Figure S13 by taking the Pearson correlation of all ΔEL/EL frames relative to the frame at 7.0 mT.

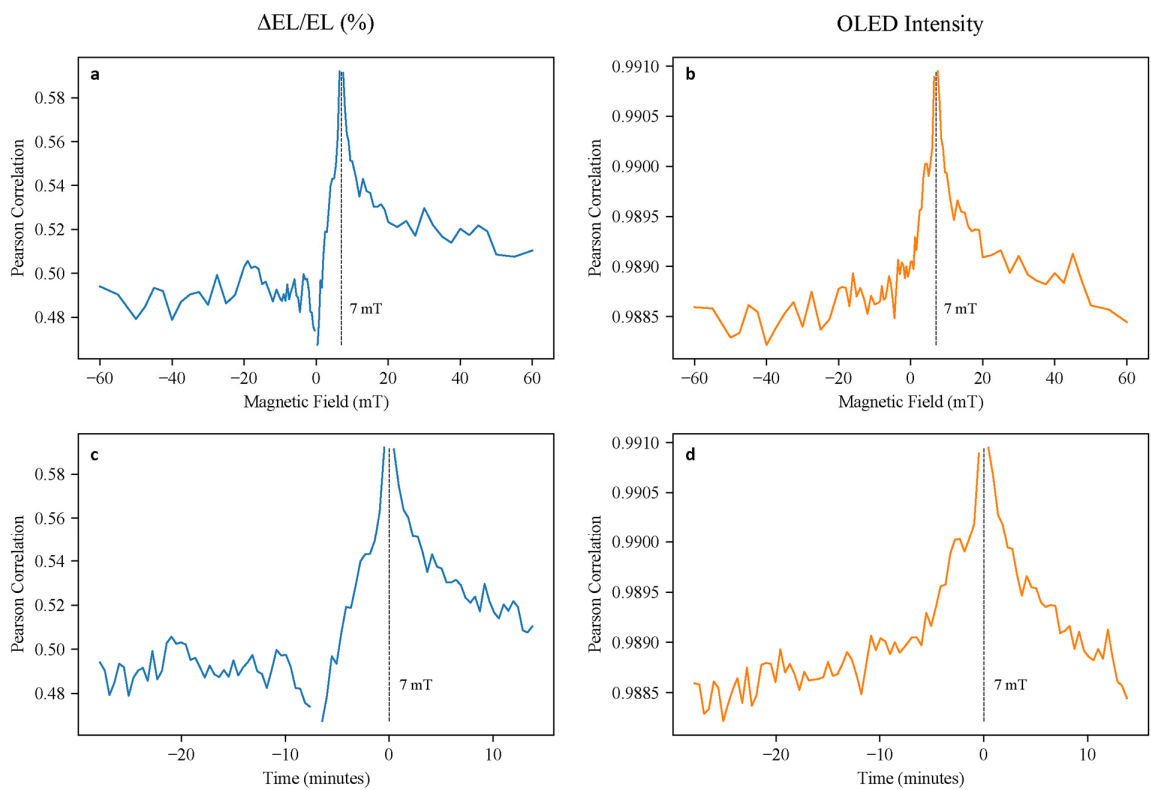

**Figure S13**: The Pearson correlation between relative (ΔEL/EL) and absolute EL intensities compared to those at 7.0 mT (indicated by the dotted line whose correlation value of 1 is excluded) for all points in the spatial map. The correlation coefficients are displayed as a function of external magnetic field (a and b) and separately the measurement time between field values (c and d). A higher degree of positive correlation is exhibited for frames surrounding 7.0 mT and a clear dependence on both the field and time can be resolved. The field dependence is manifested in (a) by an asymmetry in the positive and negative fields of equal magnitude, as well as the different rate of decay in correlation seen in time (c). The temporal dependence of the OLED intensity is heavily influenced by field-independent processes such as molecular reorganisation in the organic thin film, occurring on a minutes-timescale as estimated from (d).

**Parameter Map Evolution**

We now investigate the correlation between each half of the full magnetic field sweep which spans over both positive and negative polarities. In juxtaposition to monolithic measurements, spatially resolved MEL curves are not necessarily symmetric around the zero-field position in the magnetic field sweep, and regions of similar asymmetry occur at similar locations in the spatial parameter maps. As demonstrated in Figure S9, the clustering of high and low parameter values is not associated with the fit qualities to data of similar signal-to-noise ratios. Instead, this evolution points to the dynamics of an underlying physical process which acts to modify the magnetically-enhanced EL between polarities of the field sweep. Interestingly, this leads to parameter maps exhibiting regions of clustering at different locations across the spatial maps for negative and positive component MEL fits as shown in Figure S14. Importantly however, the correlation lengths (which measure the characteristic cluster size), as well as the parameter distributions, remain largely unchanged (Figure S15). This implies that one or more processes likely act to modify local Overhauser field strengths on a timescale of minutes, while the mean value of this quantum parameter is maintained over a large region of space. While avoid attributing any particular physical mechanism to the temporal dependence of MEL, we note that molecular and nuclear spin reorganisations occur on these timescales[2,3] and likely make significant contributions to underlying dynamics.

Furthermore, we consider the correlation of each parameter map between positive and negative field components of MEL and compare these for two separate measurements periods (15 minutes vs. 45 minutes). Considerably higher correlations for both parameters are measured at faster sweep rates for MEL under the smallest electrical straining of 0.42 mA/cm$^2$ (Figure S16). The temporal evolution of spatially correlated MEL therefore plays a significant role in quantifying the intra-device variability of spin properties in OLEDs.

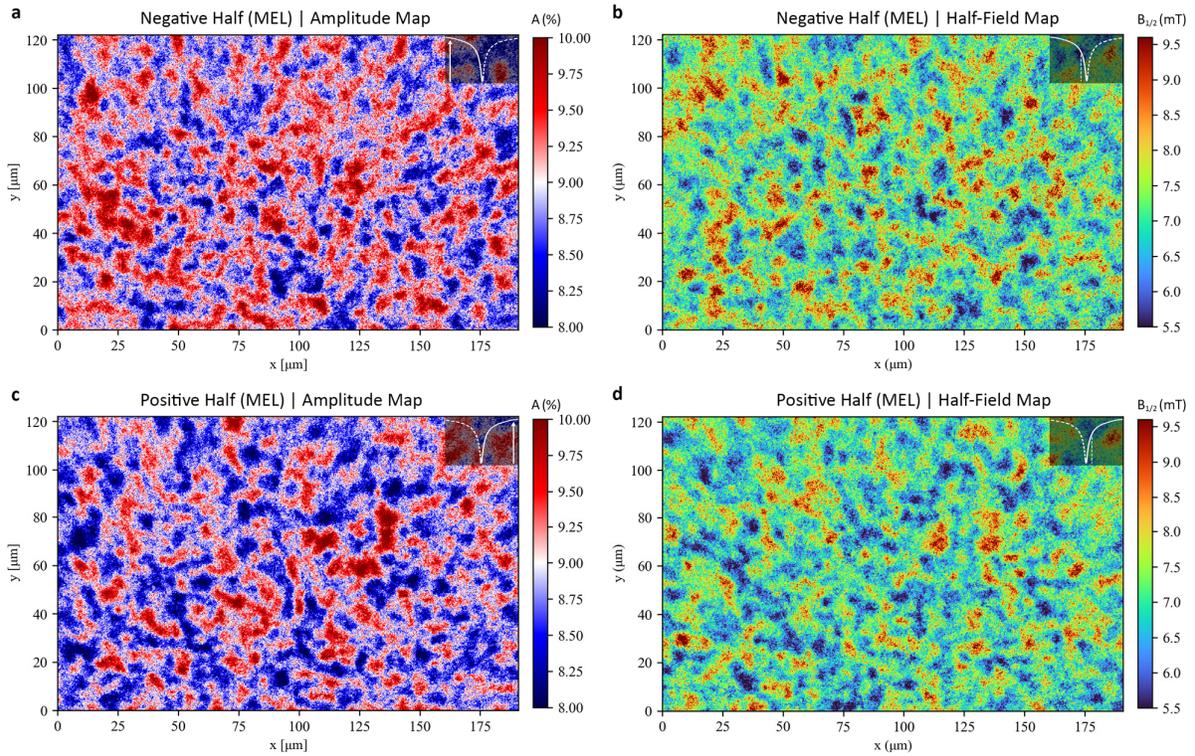

**Figure S14**: Temporal evolution of MEL in a SY-PPV OLED at 0.42 mA/cm$^2$. (a, b) The negative field component of MEL fit parameters (A, $B_{1/2}$), respectively. Compared to fits to the positive component (c, d), the spatial distributions of parameters vary substantially. Nonetheless, parameter distributions and correlograms are almost identical (Figure S15), demonstrating how substantial intra-device variation of these quantum properties remains hidden when probing the device monolithically.

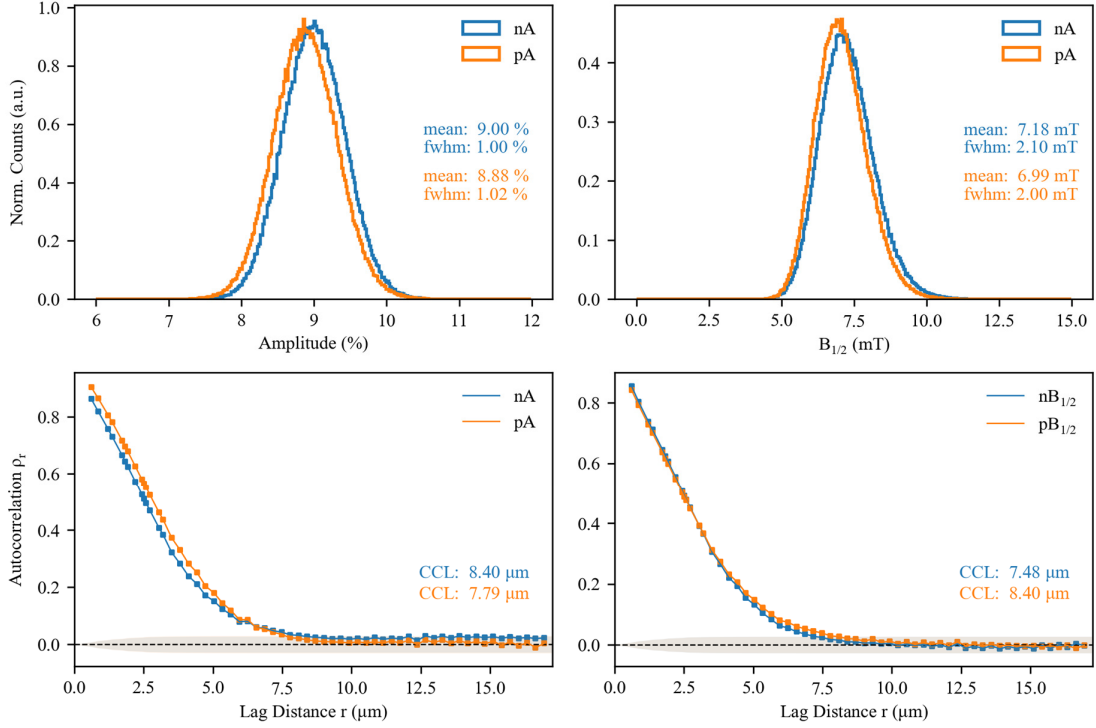

**Figure S15**: Comparison of the independently fit positive and negative field component MEL captured over a single magnetic field sweep. (a, b) Parameter distributions for A and $B_{1/2}$ resolving the intra-device variabilities for MEL parameters. We note that the OLED EL exhibits a small decay over the course of the measurement, leading to a slight shift in the mean of the amplitude (and subsequently $B_{1/2}$) distribution between the first (negative) and second (positive) components of the field sweep. (c, d) The negative and positive field components of the A and $B_{1/2}$ correlograms exhibit highly reproducible lineshapes with similar statistically significant correlation lengths.

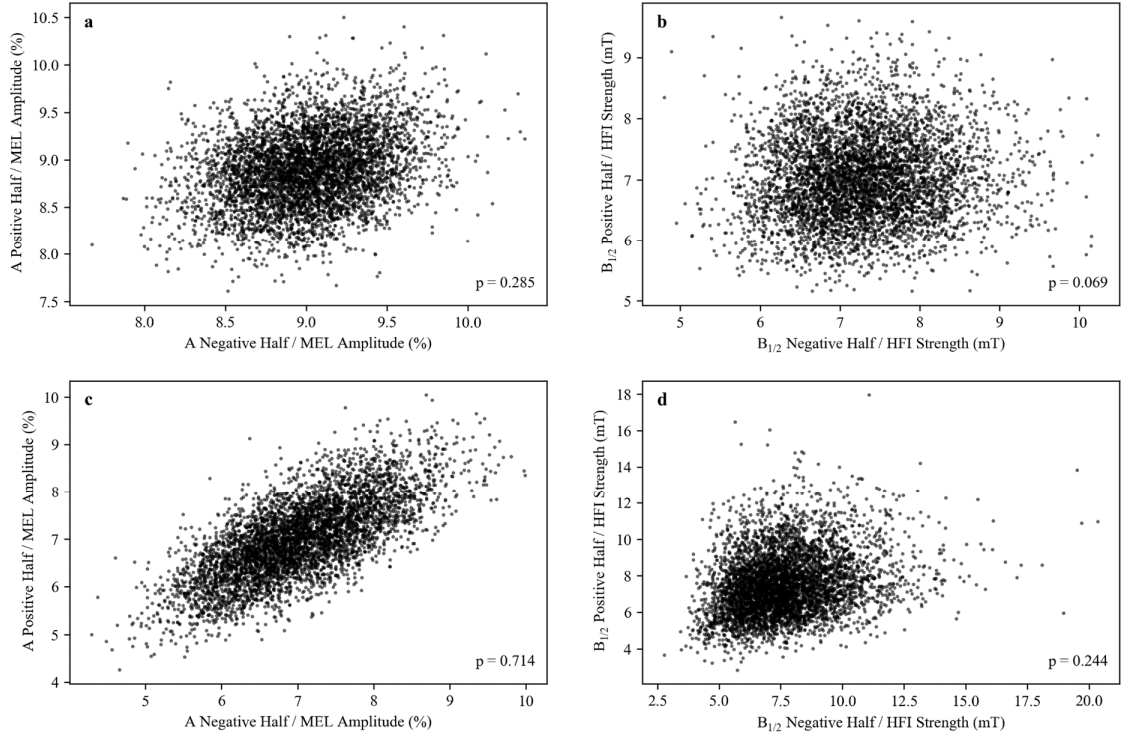

**Figure S16**: Comparison of the correlation of MEL parameter maps for positive and negative field component MEL fits at two different sweep rates. The (a, c) amplitude and (b, d) half-width correlations between spatial parameter maps over the two field polarities for the slow (45 minutes) and fast (15 minutes) measurement sweeps, respectively. A clear elevation in the correlation is observed for the faster sweep suggesting a temporal component exists in the intra-device variation of MEL.

## Intra-Device Parameter Correlations

Few investigations into the relationship between the hyperfine interaction strength ($\propto B_{1/2}$) and the saturation of the magnetically-induced change in EL (A) have been pursued, with existing studies yielding inconclusive results[4]. A major limitation in approaches taken thus far is the necessity to fabricate several devices separately from which to draw statistics from. An OLED's characteristics depend heavily upon the fabrication procedure and are sensitive to changes in environmental conditions (such as humidity), often resulting in variations to device performance which are difficult to control for. We circumvent this challenge by correlating MEL parameters spatially within a single device and for a single measurement, which also minimises the effects of material degradation on spin properties of the device. We utilise our magneto-optical setup to compare high quality fits at 10,000 regions (each 1.7 μm$^2$) across the OLED for several current densities, revealing a weak but significant positive correlation between the saturation of the MEL effect (A) and the hyperfine interaction (HFI) strength ($\propto B_{1/2}$).

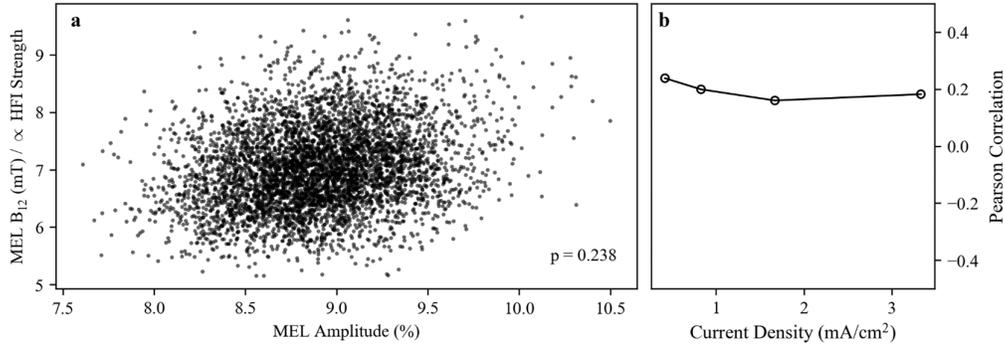

**Figure S17**: (a) A Pearson cross-correlation scatter plot between the MEL magnitude (A) and the MEL width ($B_{1/2} \propto$ hyperfine interaction strength) at 0.42 mA/cm$^2$. The amount of scatter points displayed is only 1/4 of the full dataset to assist with visual acuity, however correlation calculations are performed on the entire 10,000 points (bin 4) and exhibit fit qualities $R^2 \sim 0.99$. (b) Correlation coefficients remain positive and significant ($> 0.16$ with a 2-sided p-test $< 10^{-5}$) across all current densities.

## 5 Spatial Autocorrelation

The theory of spatial autocorrelation is used to measure any non-random pattern of attribute values over a set of spatial units. A deviation from zero spatial autocorrelation arises when observed attributes exhibit spatial patterns different to what would be expected under a random process operating in space. Positive (negative) correlations indicate a spatial similarity (dissimilarity). We implement a method of global spatial autocorrelation (Moran's I) to our spatially resolved MEL parameter maps (A, $B_{1/2}$) to quantify the behaviour in which these parameters are correlated over neighbouring spatial units. Moran's I is calculated for each lag distance as outlined in the Methods section. The strong positive autocorrelation observed below several micrometers tells us that adjacently located parameters are similarly valued (clustered) in several regions across the device (see parameter maps).

**Correlograms**

The spatial autocorrelation displayed as a function of lag distance is a useful method to estimate the typical lengths over which the MEL parameters are correlated. This occurs when autocorrelation values approach zero – that is, when the conditions of spatial randomness are satisfied. To extract a correlation length however, the statistical significance of each autocorrelation value must also be calculated to determine whether the null hypothesis can be rejected. This is done by calculating the z-score and p-value for Moran's I under the normalisation condition at each spatial lag (A and $B_{1/2}$ are normally distributed – see histograms). We use these values to generate confidence intervals of 95% and utilise the resulting confidence band as the limit to determine the statistically significant correlation lengths of the A and $B_{1/2}$ parameters.

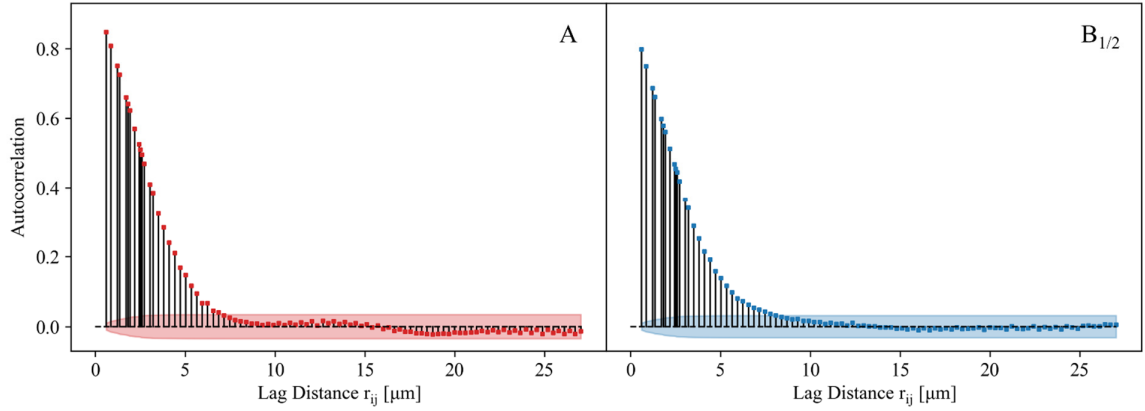

**Figure S18:** Spatial autocorrelations displayed as a function of lag distance (correlograms) of the SY-PPV OLED at 0.42 mA/cm² for parameters A (red) and $B_{1/2}$ (blue). Confidence bands using intervals of 95% are generated using z-scores and indicate regions of statistical uncertainty. The statistically significant correlation length is defined as the points which correlograms pass this band.

**Confidence Bands**

We employ a moving average (MA) model to quantify the uncertainty surrounding measures of spatial autocorrelation. Confidence bands are constructed using Bartlett's formula for variance[5]:

$$\widehat{\rho_{r=1}} = \pm z_{crit} \frac{1}{\sqrt{n}} \quad \text{and} \quad \widehat{\rho_{r>1}} = \pm z_{crit} \sqrt{\frac{1+2\sum_{i=1}^{r-1}\rho_i^2}{n}}$$

where the critical z-score ($z_{crit}$) is 1.96 for confidence intervals of 95%, and the standard error is given by $\sqrt{1/n}$ for r = 1 and $\sqrt{W_{rr}/n}$ with $W_{rr} = 1 + 2\sum_{i=1}^{r-1}\rho_i^2$ representing the estimated variance at each spatial lag for r > 1. For autocorrelation points $\rho_r$ which lie outside of these confidence bands, the null hypothesis may be rejected and the autocorrelation is statistically significant.

# 6 Magnetic Resonance

Electrically and optically detected magnetic resonance (E/ODMR) spectroscopy is a powerful technique used to detect transitions in spin-pair permutation symmetries by monitoring the sample conductivity or luminosity of the OLED under resonant excitation with oscillatory driving fields (Figure S19a). Local hyperfine disorder leads to inhomogeneous broadening around the resonance transition due to variations in the effective magnetic fields experienced by electron and hole polarons comprising the spin-pairs. Magnetic resonance signals are fit to a double Gaussian, where each polaron species contributes to the signal separately and reflects normal distributions in Overhauser fields. We assign the broader component of the magnetic resonance signals to the electron ensemble[6], while the narrower component arises from the holes (see fits in Figure S19c and S19d). This is in close agreement with recent magnetic resonance studies performed in similar devices[7,8]. These distributions provide an independent and direct measure of the average Overhauser field distributions felt by charge carriers which we compare to the half-widths of the MEL curves (and are broadened from the same underlying hyperfine interaction). The EDMR lineshapes in Figure S19b are performed at the same current densities and on the same device used in the spatially resolved MEL study, showing no change in the linewidth with operating bias in agreement with the means of $B_{1/2}$ distributions.

Similar lineshapes for both electrical and optical detection methods were found, with the advantage that EDMR produces substantially better signals compared to its optical counterpart. Optical measurements were monolithic and performed using lock-in detection methods, leading to the same loss in spatial information intrinsic to electrical readout methods. This challenge will be addressed in future work.

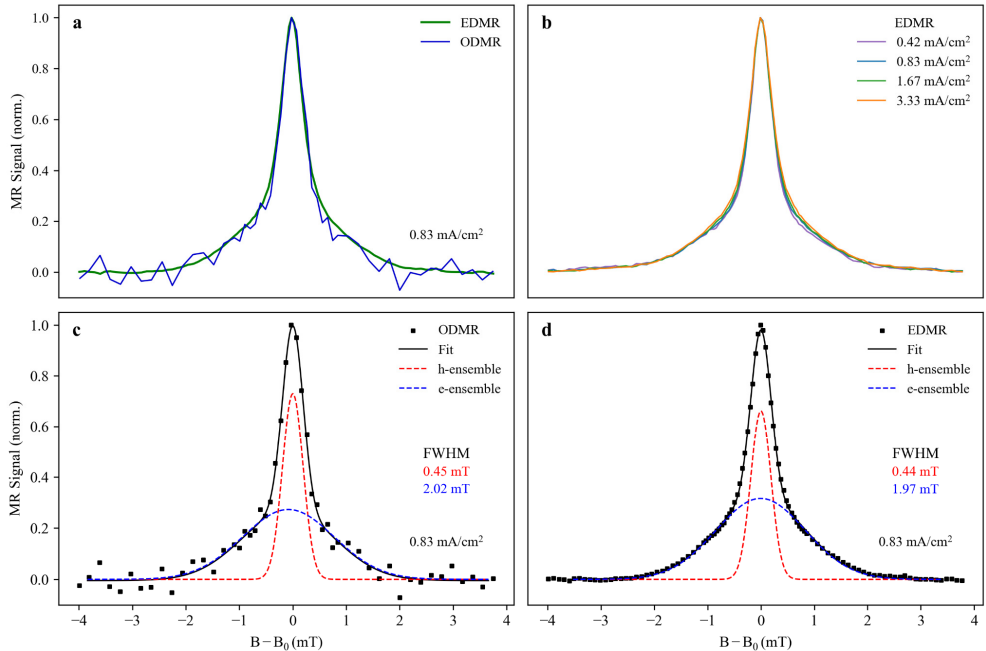

**Figure S19**: (a) Electrically and optically detected magnetic resonance (E/ODMR) of the SY-PPV OLED used for spatial resolution of MEL under a current density of 0.83 mA/cm$^2$. This current density was chosen as it produced the largest resonance signal. Electrical and optical resonances result in similar lineshapes, verifying that the Overhauser field distributions may be discerned from either of the two detection schemes. (b) EDMR for the same current densities used in the MEL study, independently verifying that the mean Overhauser field strength remains unaffected by a change in organic layer charge density. (c, d) Double Gaussian fits to the EDMR and ODMR signals, respectively. Linewidths are in close agreement for both detection schemes.

# References

1. Bayat, K., Choy, J., Farrokh Baroughi, M., Meesala, S. & Loncar, M. Efficient, uniform, and large area microwave magnetic coupling to NV centers in diamond using double split-ring resonators. *Nano Lett.* **14**, 1208–1213 (2014).

2. Liscio, F. *et al.* Molecular reorganization in organic field-effect transistors and its effect on two-dimensional charge transport pathways. *ACS Nano* **7**, 1257–1264 (2013).

3. Jiang, Y. *et al.* Direct observation of nuclear reorganization driven by ultrafast spin transitions. *Nat. Commun.* **11**, (2020).

4. Geng, R., Pham, M. T., Luong, H. M., Short, A. & Nguyen, T. D. Correlation between the width and the magnitude of magnetoconductance response in π-conjugated polymer-based diodes. *J. Photonics Energy* **8**, 1 (2018).

5. Bartlett, M. S. *The Statistical Analysis of Spatial Pattern. The Statistical Analysis of Spatial Pattern* (Springer Netherlands, 1976). doi:10.1007/978-94-009-5755-8

6. Baker, W. J., McCamey, D. R., Van Schooten, K. J., Lupton, J. M. & Boehme, C. Differentiation between polaron-pair and triplet-exciton polaron spin-dependent mechanisms in organic light-emitting diodes by coherent spin beating. *Phys. Rev. B - Condens. Matter Mater. Phys.* **84**, 1–7 (2011).

7. Grünbaum, T. *et al.* OLEDs as models for bird magnetoception: Detecting electron spin resonance in geomagnetic fields. *Faraday Discuss.* **221**, 92–109 (2019).

8. Jamali, S., Joshi, G., Malissa, H., Lupton, J. M. & Boehme, C. Monolithic OLED-Microwire Devices for Ultrastrong Magnetic Resonant Excitation. *Nano Lett.* **17**, 4648–4653 (2017).